\documentclass[amsmath,amssymb,superscriptaddress,longbibliography,twocolumn]{revtex4-1}
\usepackage{graphicx}
\usepackage{varioref}
\usepackage{xr-hyper}
\usepackage{xcolor}
\usepackage{nicefrac}
\usepackage{xfrac}
\usepackage{hyperref}
\hypersetup{colorlinks,linkcolor=red,urlcolor=purple,citecolor=orange}
\usepackage{ulem}
\usepackage{siunitx}

\newcommand{\LSCO}{La$_{2-\mathrm{x}}$Sr$_\mathrm{x}$CuO$_4$}

\newcommand{\beginsupplement}{
        \setcounter{table}{0}
        \setcounter{figure}{0}
        \renewcommand{\figurename}{\textbf{Supplementary Figure}}
        \renewcommand{\tablename}{\textbf{Supplementary Table}}
        \renewcommand{\thetable}{\arabic{table}}
}

\begin{document}

\title{Strain-Engineering Mott-Insulating La$_2$CuO$_4$}
\author{O.~Ivashko}
\email{oleh.ivashko@physik.uzh.ch}
\affiliation{Physik-Institut, Universit\"{a}t Z\"{u}rich, Winterthurerstrasse 190, CH-8057 Z\"{u}rich, Switzerland}

\author{M.~Horio}
\affiliation{Physik-Institut, Universit\"{a}t Z\"{u}rich, Winterthurerstrasse 190, CH-8057 Z\"{u}rich, Switzerland}

\author{W.~Wan}
\affiliation{Department of Physics, Technical University of Denmark, DK-2800 Kongens Lyngby, Denmark}

\author{N.~B.~Christensen}
\affiliation{Department of Physics, Technical University of Denmark, DK-2800 Kongens Lyngby, Denmark}

\author{D.~E.~McNally}
\affiliation{Swiss Light Source, Paul Scherrer Institut, CH-5232 Villigen PSI, Switzerland}

\author{E.~Paris}
\affiliation{Swiss Light Source, Paul Scherrer Institut, CH-5232 Villigen PSI, Switzerland}

\author{Y.~Tseng}
\affiliation{Swiss Light Source, Paul Scherrer Institut, CH-5232 Villigen PSI, Switzerland}

\author{N.~E.~Shaik}
\affiliation{Institute of Physics, \'{E}cole Polytechnique Fed\'{e}rale de Lausanne (EPFL), CH-1015 Lausanne, Switzerland}

\author{H.~M.~R{\o}nnow}
\affiliation{Institute of Physics, \'{E}cole Polytechnique Fed\'{e}rale de Lausanne (EPFL), CH-1015 Lausanne, Switzerland}

\author{H.~I.~Wei}
\affiliation{Department of Physics, Laboratory of Atomic and Solid State Physics, Cornell University, Ithaca, New York 14853, USA}

\author{C.~Adamo}
\affiliation{Department of Applied Physics, Stanford University, Stanford, CA 94305, USA}

\author{C.~Lichtensteiger}
\affiliation{Department of Quantum Matter Physics, University of Geneva, 24 Quai Ernest Ansermet, 1211 Geneva, Switzerland}

\author{M.~Gibert}
\affiliation{Physik-Institut, Universit\"{a}t Z\"{u}rich, Winterthurerstrasse 190, CH-8057 Z\"{u}rich, Switzerland}

\author{M.~R.~Beasley}
\affiliation{Department of Applied Physics, Stanford University, Stanford, CA 94305, USA}

\author{K.~M.~Shen}
\affiliation{Department of Physics, Laboratory of Atomic and Solid State Physics, Cornell University, Ithaca, New York 14853, USA}

\author{J.~M.~Tomczak}
\affiliation{Institute of Solid State Physics, Vienna University of Technology, A-1040 Vienna, Austria}

\author{T.~Schmitt}
\affiliation{Swiss Light Source, Paul Scherrer Institut, CH-5232 Villigen PSI, Switzerland}

\author{J.~Chang}
\email{johan.chang@physik.uzh.ch}
\affiliation{Physik-Institut, Universit\"{a}t Z\"{u}rich, Winterthurerstrasse 190, CH-8057 Z\"{u}rich, Switzerland}

\maketitle

\textbf{
The transition temperature $T_\textrm{c}$ of unconventional superconductivity is often tunable.
For a monolayer of FeSe, for example, the sweet spot is uniquely bound to titanium-oxide substrates. By contrast for La$_{2-\mathrm{x}}$Sr$_\mathrm{x}$CuO$_4${} thin films, such substrates are sub-optimal and the highest $T_\textrm{c}$ is instead obtained using LaSrAlO$_4$. 
An outstanding challenge is thus to understand the optimal conditions for superconductivity in thin films: which microscopic parameters drive the change in $T_\mathrm{c}$ and how can we tune them?
Here we demonstrate, by a combination of x-ray absorption and resonant inelastic x-ray scattering spectroscopy, how the Coulomb and magnetic-exchange interaction of La$_2$CuO$_4${} thin films can be enhanced by compressive strain. Our experiments and theoretical calculations establish that the substrate producing the largest $T_\textrm{c}$ under doping also generates the largest nearest neighbour hopping integral, Coulomb and magnetic-exchange interaction. We hence suggest optimising the parent Mott state as a  strategy for enhancing the superconducting transition temperature in cuprates.}\\[4mm]

Exposed to pressure, the lattice parameters of a material generally shrink. In turn, the electronic nearest neighbour hopping integral $t$ increases, due to larger orbital overlap. In a Mott insulator this enhancement can trigger a bandwidth-controlled insulator-to-metal transition~\cite{ImadaRMP1998}. Indeed, the ratio, $U/t$,  of the electron-electron (Coulomb) interaction $U$ and the hopping $t$ may be driven below its critical value. This premise has led to prediction of a pressure-induced insulator-to-metal transition in hypothetical solid hydrogen~\cite{Wigner35}. Experimentally, pressure-induced metallisations have been realised, e.g., in NiS$_2$~\cite{FriedemannSciRep16} and organic salts~\cite{GatieScienceAdv16}. However, besides its impact on the bandwidth, pressure also influences (in a complex fashion) the electron-electron interaction $U$ -- an effect that has received little attention so far. The fate of Mott insulators exposed to external pressure therefore remains an interesting (and unresolved) problem to consider.

In the case of layered copper-oxide materials (cuprates), superconductivity 
emerges once the Mott insulating state is doped away from 
half-filling~\cite{LeeRMP06}. In fact, it is commonly believed that the Mott 
state is a precondition for
cuprate high-temperature 
superconductivity. While the optimal doping has been established for all known 
cuprate systems, the ideal configuration -- for superconductivity -- of the 
parent Mott state has not been identified.
Typically, it is reported that hydrostatic 
pressure has a positive effect on $T_\mathrm{c}$~\cite{HardyPRL2010,WangPRB2014}. However, the 
microscopic origin of this finding remains elusive. In particular, how pressure 
influences the local Coulomb interaction $U$ and the inter-site magnetic-exchange interaction -- to lowest order -- $J_\mathrm{eff} = 4t^2/U$, is an unresolved problem.

\begin{figure*}
	\begin{center}
		\includegraphics[width=0.995\textwidth]{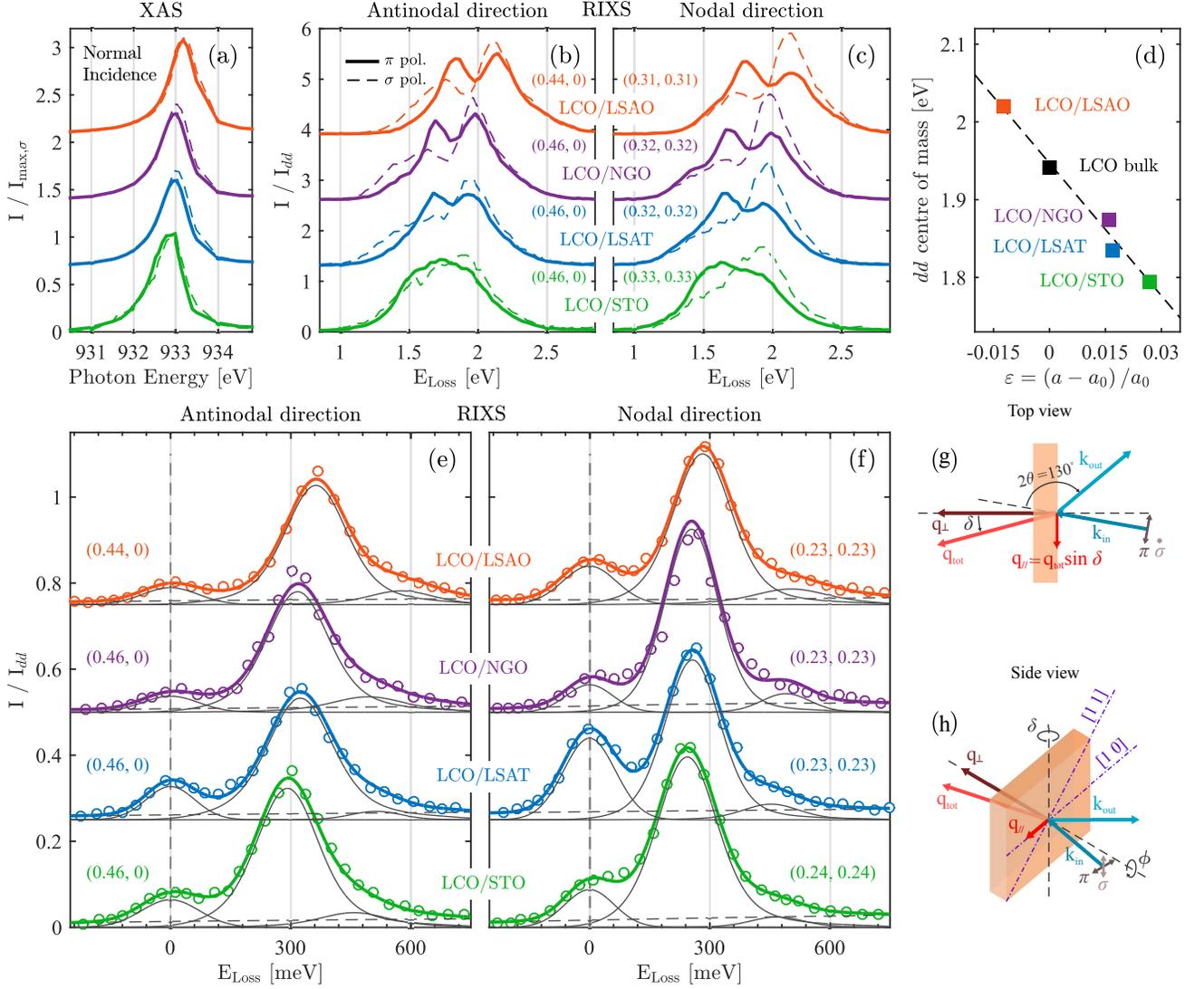}
	\end{center}
		\caption{\textbf{Strain-dependent XAS and RIXS spectra of La$_2$CuO$_4${} films.} (a) Normal incidence ($\delta=\ang{25}$) XAS spectra recorded around the copper $L_3$-edge on La$_2$CuO$_4${} films on different substrates as indicated. (b)-(f) display grazing incident copper $L_3$-edge RIXS spectra. In (b) and (c), the \textit{dd} excitations are shown for different momenta as indicated. For (a)-(c), solid (dashed) lines indicate use of $\pi$ ($\sigma$) polarised incident light. (d) displays the ``centre of mass'' of the \textit{dd} excitations vs. strain $\varepsilon$, for samples as indicated. Each (thin film) point is an average ``centre of mass'' value of all the spectra in (b) and (c). The bulk La$_2$CuO$_4${} value is extracted from Ref.~\cite{MorettiNJP11}. (e) - (f) present the low-energy part of RIXS spectra (circular points) with four-component (grey lines) line-shape fits (see text). Notice that the different film systems have, naturally, different elastic components. For visibility all curves in (a) - (e) have been given an arbitrary vertical shift. (g) and (h) illustrate schematically the scattering geometry. Source data are provided as a Source Data file.}\label{fig:fig1}
\end{figure*}

Here we present a combined x-ray absorption spectroscopy (XAS) and resonant inelastic x-ray scattering (RIXS) study of the La$_2$CuO$_4${} Mott insulating phase. We show that by straining thin films, the crystal field environment as well as the energy scales $t$ and $U$ that define the degree of electronic correlations, can be tuned.
In stark contrast to predictions for elementary hydrogen~\cite{Wigner35} and observations on standard Mott insulating compounds~\cite{FriedemannSciRep16,GatieScienceAdv16}, we demonstrate that $U/t$ remains approximately constant with in-plane strain.
 In La$_2$CuO$_4${}, both $U$ and $t$ are increasing 
with compressive strain. In-plane strain is therefore not pushing La$_2$CuO$_4${} closer to the metallisation limit. Instead, strain enhances the stiffness, \textit{i.e.} the exchange interaction $J_\mathrm{eff}$, of the antiferromagnetic ordering. These experimental observations are consistent with our band structure and constrained Random Phase Approximation (cRPA) calculations that reveal the same trends for
$t$, $U$ and $J_\mathrm{eff}$. For superconductivity, originating from the anti-ferromagnetic pairing channel, the exchange interaction is a key energy scale. Our study demonstrates how $J_\mathrm{eff}$ can be controlled and enhanced with direct implications 
for the optimisation of superconductivity.\\

\begin{figure*}
 	\begin{center}
 		\includegraphics[width=0.995\textwidth]{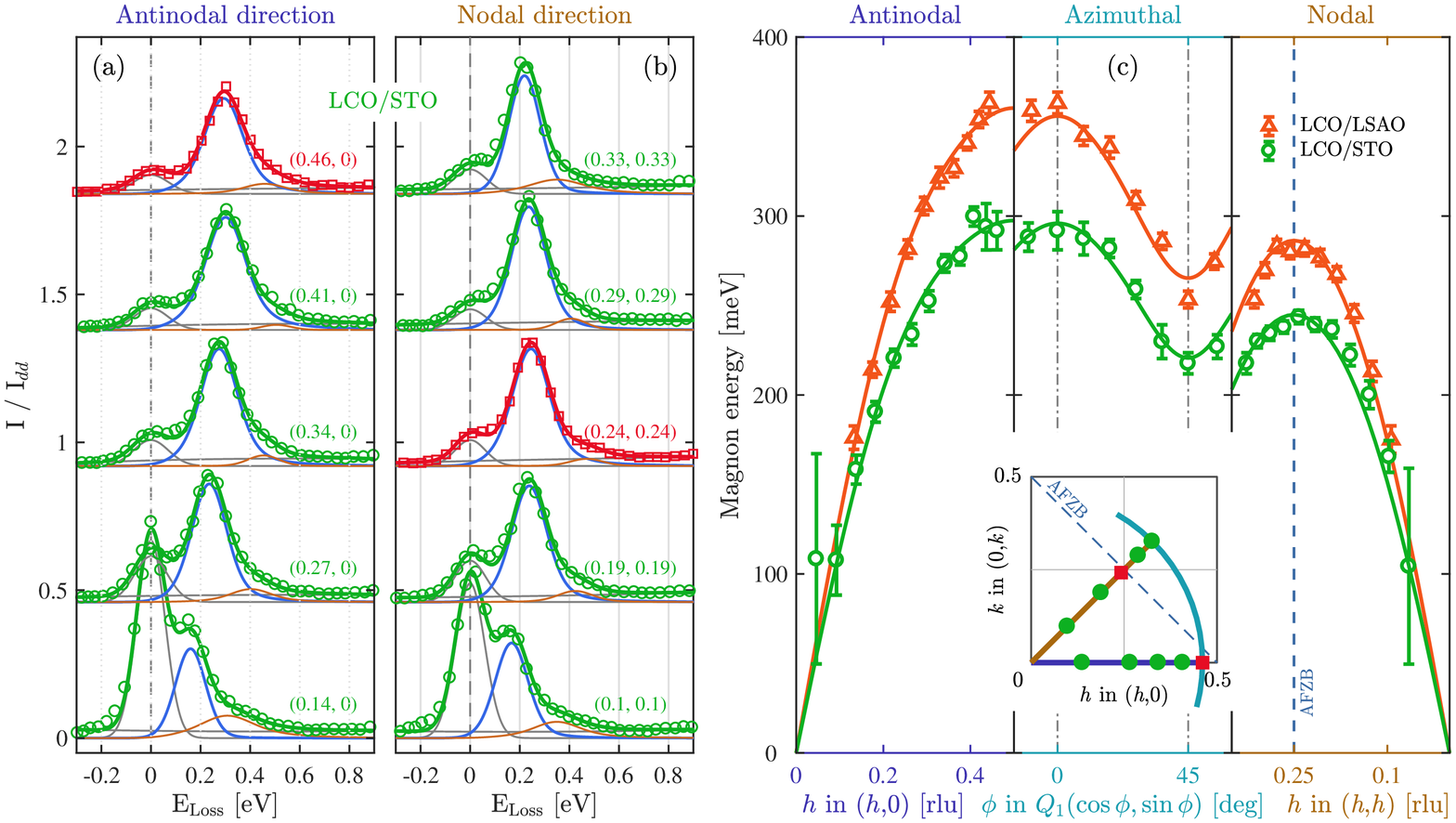}
 	\end{center}
 	\caption{\textbf{Magnon Dispersions of La$_2$CuO$_4${} thin films.} (a) and (b) display raw RIXS spectra recorded on the LCO$\slash$STO  thin film system, along the antinodal [1 0] and nodal [1 1] directions, respectively. Red curves represent the data close to the antiferromagnetic zone boundary (AFZB) as shown in the inset in (c). Solid lines are  fits to the data (see text for detailed description). Notice that elastic scattering is, as expected, enhanced as the specular condition $(0,0)$ is approached. In (c) magnon dispersions of LCO$\slash$LSAO and LCO$\slash$STO, extracted from fits to raw spectra as in (a) and (b), along three different momentum trajectories (see solid lines in the inset) are presented. Solid lines through the data points are obtained from two-dimensional fits using Hubbard model (see the main text). Error bars are three times the standard deviations  extracted from the fits. In (c) $Q_1$ takes different values for each compound due to slightly different incident energies and in-plane lattice parameters, resulting in $0.4437$ ($0.4611$) for LCO$\slash$LSAO (LCO$\slash$STO). Source data are provided as a Source Data file.}\label{fig:fig2}
\end{figure*}

\begin{figure*}
	\begin{center}
		\includegraphics[width=0.99\textwidth]{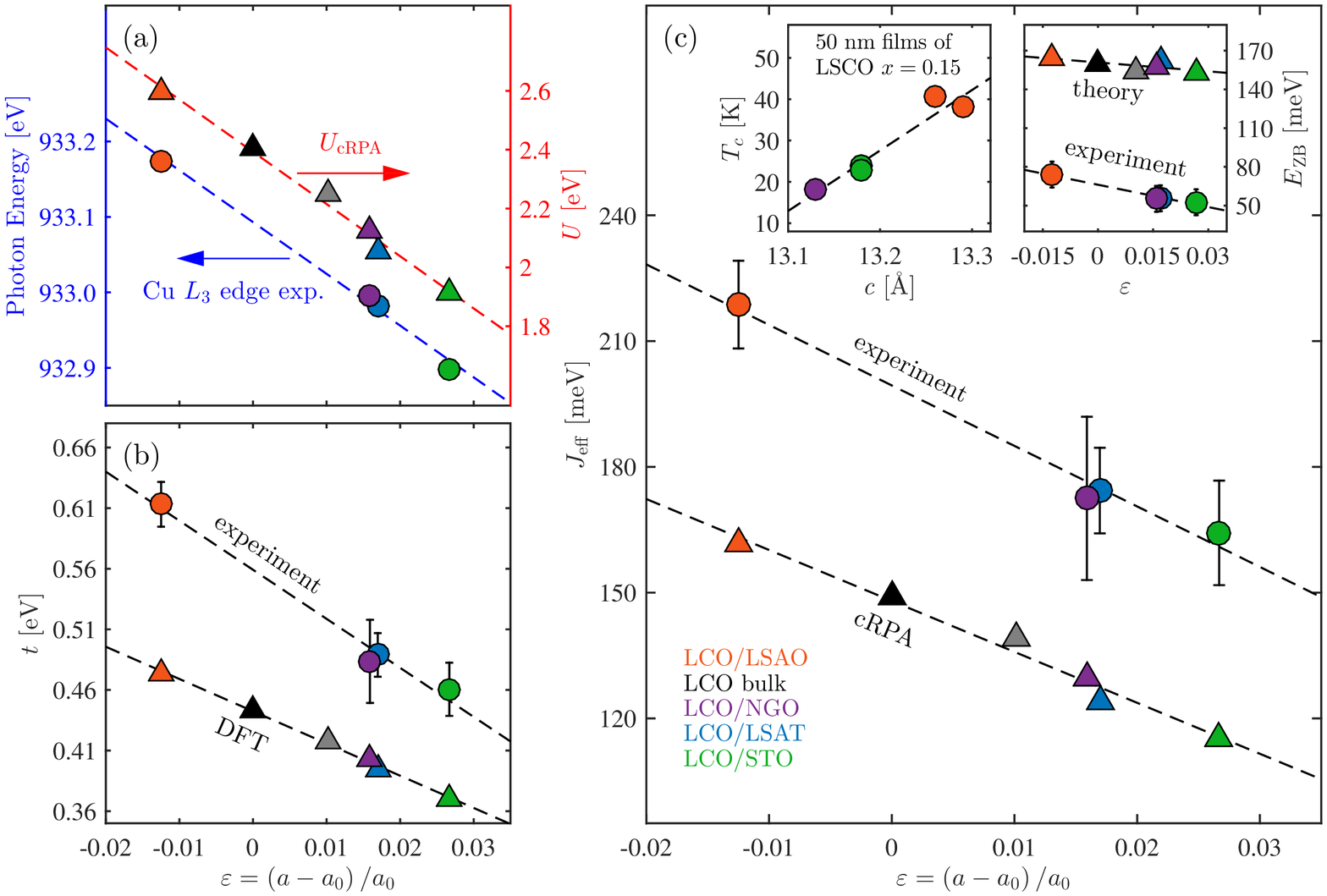}
    \end{center}
        \caption{\textbf{Cuprate energy scales versus strain.}  In (a) XAS at Cu $L_3$-edge resonances (left) and  the theoretical results for Coulomb interaction $U$ (right) are presented. Experimental and theoretical derived hopping parameters $t$ are presented in (b), as indicated. Notice that $t$ is not scaling with the copper-oxygen bond length $r^{-\alpha}$ with $\alpha=6-7$ as sometimes assumed~\cite{HafligerPRB2014}. $J_\mathrm{eff}$ as a function of $\varepsilon = \left( a- a_0 \right) / a_0$ (where $a_0$ is the in-plane bulk lattice parameter) is presented in (c) for both theoretical and experimental results. Zone-boundary dispersion $E_\mathrm{ZB}$, extracted from the Hubbard model, for experimental and theoretical parameters, are presented in the right inset of (c) as a function of strain $\varepsilon$. Superconducting transition temperature $T_\mathrm{c}$ as a function of out-of-plane lattice constant \textit{c} -- for optimally doped LSCO thin films (see Supplementary Table~1
        -- is presented in the left inset in (c). Colour code for the data points in the figure refers to the one shown in (c). The error bars for the experimental data are standard deviations extracted from the fits. The theoretical value corresponding to $\varepsilon \approx 0.01$ (gray symbols) is an artificial sample as described in Table~\ref{tab:tab1}. Source data are provided as a Source Data file.}\label{fig:fig3}
\end{figure*}

\textbf{Results}\\
\textbf{Crystal-Field Environment:} 
Thin films ($8-19$~nm) of La$_2$CuO$_4$ grown on  
substrates with different lattice parameters are studied. 
In this fashion both compressive [LaSrAlO$_4$ (LSAO)] and tensile [NdGaO$_3$ (NGO), (LaAlO$_3$)$_{0.3}$(Sr$_2$TaAlO$_6$)$_{0.7}$ (LSAT) and SrTiO$_3$ (STO)] strain is imposed.
Strain is defined by $\varepsilon\equiv (a-a_0)/a_0$ where $a$ and $a_0$ are the in-plane lattice parameters of the thin film
and the bulk, respectively. 
For the above-mentioned samples, $\varepsilon=$ -1.25, 1.59, 1.70 and 2.67\% is obtained respectively.
The substrates are tuning both the in- and out-of-plane lattice 
parameters of the La$_2$CuO$_4${} films (see Table~\ref{tab:tab1}), directly affecting the electronic energy scales of the system. This can be readily observed from the XAS spectra at the copper $L_3$ edge [see 
Fig.~\ref{fig:fig1}(a)]. A considerable shift ($\sim280$~meV) of the Cu $L_3$ 
edge is found when comparing the compressive strained LCO$\slash$LSAO with 
the tensile strained LCO$\slash$STO film. 
The \textit{dd} excitations 
probed through the Cu $L_3$ edge, exhibit a similar systematic shift [see 
Fig.~\ref{fig:fig1}(b-d)]. The strain-dependent line shape of the \textit{dd} excitations, points to a change in the crystal-field environment.
The double peak structure, known for bulk La$_2$CuO$_4${}~\cite{MorettiNJP11,PengNatPhys2016}, is also found in our thin films on  LSAO, NGO and LSAT substrates.
For LCO$\slash$STO, however, a more featureless line shape is found, 
resembling doped La$_{2-\mathrm{x}}$Sr$_\mathrm{x}$CuO$_4${} (LSCO)~\cite{IvashkoPRB2017}. All together, the shift of the Cu $L_3$ edge and the centre of 
mass [Fig.\ref{fig:fig1}(d)] of the \textit{dd} excitations (along with the 
line-shape evolution) demonstrate the effectiveness of epitaxial strain for tuning the electronic excitations.\\

\textbf{Zone-boundary Magnons:}
The low-energy part of the RIXS spectra in the vicinity to the high-symmetry 
zone-boundary points $(\nicefrac{1}{2},0)$ and 
$(\nicefrac{1}{4},\nicefrac{1}{4})$ are shown in Fig.~\ref{fig:fig1}(e,f). 
Generally, the spectra are composed of elastic scattering, a magnon and a weaker 
multimagnon contribution on a weak smoothly-varying background. In all 
zone-boundary (ZB) spectra, the magnon excitation is by far the most intense 
feature. The ZB magnon excitation energy scale, can thus be 
extracted by the naked eye. Comparing antinodal zone boundary spectra for the 
compressive (LSAO) and tensile (STO) strained systems [Fig.~\ref{fig:fig1}(e)] 
reveals a softening of about $60$~meV in the STO system. To first order, the 
magnetic-exchange interaction $2J_\mathrm{eff}$ is setting the antinodal ZB 
magnon energy scale~\cite{ColdeaPRL01,HeadingsPRL10}. Without any sophisticated analysis, we thus can conclude that the magnetic-exchange interaction of LCO thin films can be tuned by strain.  At the nodal ZB
this effect is much 
less pronounced [see Fig.~\ref{fig:fig1}(f)], suggesting a strain dependent 
zone-boundary dispersion. Therefore, the central experimental observations, 
reported here, are that the crystal-field environment, the magnetic exchange 
interaction and the magnon zone-boundary dispersions are tunable through strain.\\

\textbf{Magnon dispersion:} To extract the magnetic-exchange interaction $2J_\mathrm{eff}$ in a more quantitative 
fashion, three steps are taken. First, a dense grid of RIXS spectra has been
measured along the nodal and antinodal directions in addition to a 
constant-$|{\textbf{q}}_{/\!/}|$ trajectory connecting the two [see inset of 
Fig.~\ref{fig:fig2}(c)]. Second, fitting these spectra allows extracting the full magnon dispersion for all the  film systems. Finally, these dispersions are parametrised using strong-coupling perturbation theory for the Hubbard model to extract the effective magnetic exchange couplings.

Compilations of nodal and antinodal RIXS spectra are shown in 
Fig.~\ref{fig:fig2}(a,b) and Supplementary Fig.~1
The magnon 
excitations remain clearly visible even near the zone centre where elastic 
scattering is typically enhanced. To extract the magnon dispersion, such spectra 
were fitted using a Gaussian line shape for the elastic scattering and a 
quadratic function for the weak background. The width of the 
elastic Gaussian is
a free parameter in order to account for a 
small phonon contribution. Two antisymmetric Lorentzian 
functions~\cite{TaconNATP11,MonneyPRB16,LamsalPRB16} for the magnon and the 
small multimagnon signals were adopted and convoluted with the experimental 
resolution function. The quality of the fits can be appreciated from 
Fig.~\ref{fig:fig1}(e,f) and Fig.~\ref{fig:fig2}(a,b). Generally, the magnon 
width was found to be comparable to the experimental resolution and independent of momentum, suggesting that the line shape is resolution limited. In contrast 
to doped systems~\cite{DeanNATM13,IvashkoPRB2017}, magnons of the LCO thin 
films have a negligible damping and hence the pole of the fitted excitation coincides essentially with the peak maxima.

The extracted magnon dispersions of LCO$\slash$STO and LCO$\slash$LSAO are 
displayed in Fig.~\ref{fig:fig2}(c). The analysis confirms that the magnon 
bandwidths are significantly different for the two systems.  Near the 
$(\nicefrac{1}{2},0)$ zone-boundary point, the LCO$\slash$LSAO magnon reaches 
about 360 meV whereas for LCO$\slash$STO a comparative softening of 60~meV is 
found [Fig.~\ref{fig:fig2}(c)]. This softening is less pronounced near the 
$(\nicefrac{1}{4},\nicefrac{1}{4})$ zone-boundary point 
[Fig.~\ref{fig:fig2}(c)], demonstrating that the zone boundary dispersion is 
also strain dependent [see right inset of Fig.~\ref{fig:fig3}(c)].
Our results thus show a larger ZB dispersion for the LCO$\slash$LSAO system.\\

\textbf{Discussion}\\
Different theoretical models have been applied to analyse RIXS spectra of the cuprates. Many of these approaches are purely numerical starting either from a metallic or localised picture~\cite{NomuraJPSJ2015,JiaNATC2014}. To parameterise experimental results, analytical models are useful. The Hubbard model has, therefore, been frequently used to describe the magnon dispersion of 
La$_2$CuO$_4${}~\cite{DelannoyPRB2009,HeadingsPRL10,ColdeaPRL01,DallaPiazzaPRB12, 
IvashkoPRB2017}.
We employ a $U-t-t^\prime-t^{\prime\prime}$ single-band Hubbard model, since $t^\prime$ and $t^{\prime\prime}$ hopping integrals have previously been shown relevant to account in detail for the magnetic dispersion~\cite{DelannoyPRB2009,DallaPiazzaPRB12}. By mapping onto a Heisenberg Hamiltonian, an analytical expression (see method section) for the magnon dispersion $\omega(\textbf{q})$~\cite{DelannoyPRB2009,DallaPiazzaPRB12,IvashkoPRB2017} has been derived.

Before fitting our results, it is useful to consider the ratios $U/t$, $t^\prime/t$ and 
$t^{\prime\prime}/t^\prime$ for single-layer cuprate systems. 
In-plane strain will enhance oxygen-$p$ to copper-$d$ orbital hybridizations 
and hence the effective nearest-neighbour hopping $t$ in the one-band Hubbard model description~\cite{AbrechtPRL03}. 
This trend can be calculated from approximate numerical methods such as 
density functional theory (DFT) [see Fig.\ref{fig:fig3}(b)]. 
Besides this increase in band-width, oxygen-$p$ orbitals are concomitantly pushed down 
[Supplementary Figure~4
] and the $e_g$-splitting is expected to increase.
Indeed, as shown in Fig.~\ref{fig:fig1}(d), the RIXS $dd$ excitations (``centre of mass'') are
pushed to higher energies upon compressive strain--consistent with an enhanced $e_g$ splitting.
This tendency of states moving away from the Fermi-level is generally expected to diminish their
ability to screen the Coulomb interaction. Its local component--the Hubbard $U$--quantifies 
the energetic penalty of adding a second electron to the half-filled effective $d_{x^2-y^2}$ orbital.
To a good approximation the evolution of this process is accessible by tracking the
Cu $2p^63d^9 \rightarrow 2p^53d^{10}$ XAS resonance. As seen in Fig. 1(a) and 3(a), we find the copper $L$-edge resonance 
to shift notably upwards under in-plane compression.
This strongly suggests that the energy cost for double occupancies--and hence the Hubbard $U$--increases under compressive strain,
confirming the above rationale.

Beyond the suggested impact on screening, pressure or strain also modify the localization of the effective $d_{x^2-y^2}$ orbital.
As a basis-dependent quantity, the Hubbard $U$ is sensitive also to this second mechanism~\cite{TomczakPRB2009}.
To corroborate our experimental finding for the effective Coulomb interaction under in-plane strain,
we therefore carried out cRPA calculations for La$_2$CuO$_4$ that include both screening and basis-localization effects~\cite{TomczakPRB2009,TomczakPRB2010,SantePRB2017,KimPRB2018} (see Methods section).
We stress that cRPA is an approximate numerical approach. It is known that correlation effects are underestimated when using only the static limit of the Hubbard interaction~\cite{WernerPRB2015}. Indeed the cRPA obtained $U\approx 5t$ [see Fig.~\ref{fig:fig3}(a,b) and Table~\ref{tab:tab1}] is below the expected bandwidth-controlled threshold value.
We therefore focus on the relative trends produced by the cRPA. As shown in
Fig.~3(a), the simulation indeed predicts the Hubbard $U$ to increase with compressive strain.
This confirms the above rationale and thus enables us to interpret the XAS Cu $L_3$-edge
as a proxy for the variation of the screened Coulomb interaction $U$.
The fact that both $U$ and $t$ increase linearly with compressive strain [Fig. 3(a,b)] leads us to the Ansatz
that the ratio $U/t$ is approximately constant. In the following analysis of the experimental data, we therefore
assume $U/t = 9$ \cite{DelannoyPRB2009} (and $t^{\prime\prime}/t^\prime = -0.5$~\cite{YoshidaPRB06}).

In this fashion, our Hubbard model effectively depends only on $t$ and $t'$, that constitute our fitting parameters. On 
a square lattice, one would expect $t'/t$ to remain approximately constant as a 
function of strain. Indeed, fitting the magnon dispersions, yields that $t$ 
increases with reduced lattice parameter [Fig.~\ref{fig:fig3}(b)] while 
$t'/t\approx -0.4$ (see Table~\ref{tab:tab1}). This value of $t'/t$ is reasonably 
consistent with ARPES and LDA derived band structures of the most tetragonal 
single-layer cuprate systems Hg1201 and 
Tl2201~\cite{SakakibaraPRL10,SakakibaraPRB12,PlatePRL05,MattNatComm2018}. 
Single-band tight-binding models, applied to LSCO, have found significantly lower values of $t^\prime/t$~\cite{YoshidaPRB06,ChangNATC13,ChangPRB2008a}. However, 
when including hybridisation between the $d_{x^2-y^2}${} and $d_{z^2}${} orbitals, existing 
in LSCO, one again finds $t^\prime/t\approx 
-0.4$~\cite{MattNatComm2018,SakakibaraPRL10,SakakibaraPRB12}.
The described variation of the hopping $t$ and the Hubbard $U$ translates into a pressure-dependent magnetic exchange interaction $J_\textrm{eff}$ when mapping the Hubbard model (at strong coupling) into a Heisenberg Hamiltonian:  $J_\textrm{eff}=4t^2/U - 64t^4/U^3$ (see Eq.~\ref{eq:Jeff}). Since $U/t \approx const.$, it is therefore expected that $J_\textrm{eff}$ scales with $t$.
Indeed, as directly visible from the magnon dispersion and our cRPA calculations [Fig.~\ref{fig:fig3}(c)], $J_\mathrm{eff}$ increases linearly when going from tensile ($\epsilon>0$)  to compressive strain ($\epsilon<0$) .

In LSCO system superconductivity emerges upon hole doping.
It is known that for LSCO, the highest $T_\mathrm{c}$ is 
reached when thin films are grown on LSAO substrates~\cite{SatoPhysicaC1997,LocquetActaP1997}.
Although higher $T_\mathrm{c}$ has been linked to larger $c$-axis parameter [see left inset of Fig.~\ref{fig:fig3}(c)], the physical origin of 
this effect has remained elusive. In-plane strain also tunes the $c-$axis lattice parameter through the Poisson ratio~\cite{NakamuraPRB2000}. The observed evolution of the \textit{dd}-excitations [Fig.~\ref{fig:fig1}(d)] is consistent 
with a compressive strain-induced enhancement of the $e_g$ splitting. It has been argued that this \textit{orbital distillation} (avoidance of $d_{x^2-y^2}${} and $d_{z^2}${} hybridisation) is beneficial for superconductivity~\cite{MattNatComm2018,SakakibaraPRL10,SakakibaraPRB12}.
The $e_g$ splitting might also indirectly increase $T_\mathrm{c}$ by changing the screening of the local Coulomb interaction $U$, as described above.
Antiferromagnetic interactions are a known source 
for $d$-wave Cooper pairing~\cite{ScalapinoRMP12}. 
A link between $J_\mathrm{eff}$ and $T_\mathrm{c}$ is therefore expected in the large $U/t$ limit~\cite{OferPRB2006,EllisPRB2015,FratinoSciRep2016}. Here, we have explicitly demonstrated how the important energy scale $J_\mathrm{eff}$ can be tuned through strain. This direct connection 
between lattice parameters and the magnetic exchange interaction in Mott insulating La$_2$CuO$_4${} provides an engineering principle for the optimisation of high-temperature cuprate superconductivity. \\

Our study highlights the power of combining oxide molecular beam epitaxial material design with synchrotron spectroscopy. In this particular case of La$_2$CuO$_4$\ thin films, it is shown how Coulomb and antiferromagnetic exchange interactions can be artificially engineered by varying the film substrate. In this fashion, direct design control on the Mott insulating energy scales, constituting the starting point for high-temperature superconductivity, has been reached. It would be of great interest to apply this strain-control rationale to doped single-layer HgBa$_2$CuO$_{4+x}$ and Tl$_2$Ba$_2$CuO$_{6+x}$ cuprate superconductors to further enhance the transition temperature $T_\mathrm{c}$. \\[2mm]

\textbf{Methods}\\
{\label{methods}}
\textit{Film systems:}
High quality La$_2$CuO$_4${} (LCO) thin films were grown using Molecular 
Beam Epitaxy (MBE), on four different substrates: (001)$_\mathrm{c}-$SrTiO$_3$ 
(STO), (001)$_\mathrm{c}-$(LaAlO$_3$)$_{0.3}$(Sr$_2$TaAlO$_6$)$_{0.7}$ (LSAT), 
(001)$_\mathrm{pc}-$NdGaO$_3$ (NGO) and (001)$_\mathrm{c}-$LaSrAlO$_4$ (LSAO). 
Comparable LCO film thickness, for STO \& LSAT and for NGO \& LSAO, were used 
(Table~\ref{tab:tab1}). For such thin films, the in-plane lattice parameter 
$a_\mathrm{film}$ is set by the substrate lattice $a$ indicated in 
Table~\ref{tab:tab1}. Compared to bulk LCO, the substrate STO, LSAT and NGO 
induce tensile strain whereas LSAO generates compressive strain.
Film thicknesses were extracted from fit to the $2\theta$ scans (see Supplementary Figure~2
using an x-ray diffraction tool as in Ref.~\cite{LichtensteigerJAC2008}. \\

\textit{Spectroscopy experiments:} 
X-ray absorption spectroscopy (XAS) and resonant inelastic x-ray scattering 
(RIXS) were carried out at the ADRESS 
beamline~\cite{ghiringhelliREVSCIINS2006,strocovJSYNRAD2010} of the Swiss Light 
Source (SLS) synchrotron at the Paul Scherrer Institut. All data were collected at base temperature 
($\sim$~$20$~K) of the manipulator under ultra high vacuum (UHV) conditions,  
$10^{-9}$~mbar or better. RIXS spectra were acquired in grazing exit geometry 
with both linear horizontal ($\pi$) and linear vertical ($\sigma$) incident 
light polarisation with a scattering angle $2\mathrm{\theta}=\ang{130}$ [see Fig.~\ref{fig:fig1}(g,h)]. An 
energy resolution half width at half maximum (HWHM) of $68$~meV -- at the Cu $L_3$ edge -- was extracted from 
the elastic scattering signal. Momentum 
$\textbf{q}=\mathrm{\textbf{q}}_{/\!/}=(h,k)$ is 
expressed in reciprocal lattice units (rlu).\\

\begin{table*}[ht]
\begin{center}
\begin{ruledtabular}
\caption{\textbf{Lattice and model parameters for the different La$_2$CuO$_4${} film systems.} Thickness $h$ of the thin films (measured by x-ray diffraction) is indicated for substrates as indicated. For the films and bulk LCO, $a$ indicates the room temperature substrate and average in-plane lattice parameter, respectively. The $c$-axis lattice parameters were measured directly by x-ray diffraction (room temperature) on our films whereas for bulk LCO, the literature value is given~\cite{RadaelliPRB94}. For the ``Artificial LCO film'', $c$-axis was interpolated from the measured samples assuming an in-plane lattice parameter of $3.842$~\AA. Values of $t$ and $t^\prime$ obtained from the fit using a Hubbard model with $U/t=9$, $Z=1.219$ (quantum renormalisation factor)~\cite{DelannoyPRB2009,IvashkoPRB2017} and $t^{\prime\prime}=-t^\prime$/2. The corresponding theoretical DFT and cRPA results are also included. DFT hopping parameters were obtained using an effective single-band model. Both the screened ($U$) and bare ($v$) interaction increase with in-plane strain within the cRPA methodology. The substrate lattice parameters are taken from Refs.~\onlinecite{SatoPhysicaC1997,VasylechkoJAC2000}. Source data are provided as a Source Data file.}\label{tab:tab1}
\begin{tabular}{ccccccccccc}
Sample & $h$ [nm] & $a$ [\AA] & $c$ [\AA] & $t$  [meV] & $-t^\prime/t$  & $t$  [meV] & $-t^\prime/t$ & $-t^{\prime\prime}/t^\prime$  & $U$ [eV] & $v$ [eV]\\
&&&&exp.&exp.&DFT&DFT&DFT&cRPA&cRPA\\\hline
LCO$\slash$STO & 7$-$8 & 3.905 & 12.891 & 460.5 & 0.389 & 369.6 & 0.0908 & -0.044 & 1.92& 12.76\\
LCO$\slash$LSAT & 7$-$8 & 3.868 & 12.981 & 488.9 & 0.387 & 395.0 & 0.0907 & 0.165 & 2.05 & 13.06\\
LCO$\slash$NGO & 17$-$19 & 3.864 & 13.077 & 483.6 & 0.388 & 416.1 & 0.0910 & 0.335 & 2.12& 13.24\\
LCO$\slash$LSAO & 18$-$19 & 3.756 & 13.195 & 613.2 & 0.422 & 473.7 & 0.0917 & 0.640 & 2.60 & 14.25\\
Bulk LCO & --- & 3.803 & 13.156 & --- & --- & 443.7 & 0.0915 & 0.510 & 2.40& 13.86 \\
``Artificial LCO film'' & --- & 3.842 & 13.105 & --- & --- & 417.9 & 0.0917 & 0.361 & 2.25 & 13.54\\
\end{tabular}
\end{ruledtabular}
\end{center}	
\end{table*}

\textit{Hubbard Model:}
A single-band Hubbard model is adopted in the present study. Being important to 
consider a second-neighbour hopping integral -- for La$_2$CuO$_4${} 
compound~\cite{DelannoyPRB2009,DallaPiazzaPRB12} -- in order to fully describe the magnon 
dispersion relation~\cite{ColdeaPRL01,HeadingsPRL10} we consider the following 
Hamiltonian:
\begin{equation}
\begin{split}
    H = & - t \sum_{\left<i,j\right>,\sigma} c_{i,\sigma}^\dag c_{j,\sigma}  - t^\prime \sum_{\left<\left<i,j\right>\right>,\sigma} c_{i,\sigma}^\dag c_{j,\sigma} \\
        & - t^{\prime\prime} \sum_{\left<\left<\left<i,j\right>\right>\right>,\sigma} c_{i,\sigma}^\dag c_{j,\sigma} + U \sum_i n_{i,\uparrow} n_{i,\downarrow}
\end{split}
\end{equation}
where $t$, $t^\prime$ and $t^{\prime\prime}$ are the first-, second- and 
third-nearest-neighbour hopping integrals; $U$ is the on-site Coulomb 
interaction integral; $c_{i,\sigma}^\dag$ and $c_{i,\sigma}$ are the creation 
and annihilation operators at the site $i$ and spin 
$\sigma=\uparrow,\downarrow$; and $n_{i,\sigma} \equiv c_{i,\sigma}^\dag 
c_{i,\sigma}$ is the density operator at the site $i$ with spin $\sigma$. The 
sum (for the hopping process) is done over the first- $\left<\star\right>$, 
second- $\left<\left<\star\right>\right>$ and third-nearest neighbour sites 
$\left<\left<\left<\star\right>\right>\right>$.\\
Using this Hamiltonian at 
strong coupling it is possible to obtain~\cite{DelannoyPRB2009,DallaPiazzaPRB12} a magnon 
dispersion of the form:
\begin{equation}
    \omega(\textbf{q})=Z\sqrt{A(\textbf{q})^2-B(\textbf{q})^2}.
\end{equation}

The momentum dependence of $A$ and $B$ can be expressed in terms of trigonometric functions

\begin{align*}
P_j(h,k) & = \cos{j h a} + \cos{j k a} \\
X_j(h,k) & = \cos{j h a}~\!\cos{j k a}\\ 
X_{3a}(h,k) & = \cos{3 h a}~\!\cos{k a} + \cos{h a}~\!\cos{3 k a}
\end{align*}
such that~\cite{IvashkoPRB2017}:
\begin{equation}
\begin{split}
    A & = 2 J_1 + J_2 \left(  P_2 - 8 X_1 -26 \right) + 2 J_1^\prime \left(  X_1- 1  \right) \\
       & + \left[J_1^{\prime\prime}  - \frac{8 J_1}{U^2} \left( -t^{\prime 2} + 4 t^\prime t^{\prime\prime} - 2 t^{\prime\prime 2 } \right)\right] \left(  P_2 - 2\right) \\
       & + 2 J_2^\prime \left( - 2 P_2 + 4 X_1 + X_2 - 1 \right) \\
       & + \frac{2 J_1^\prime J_1^{\prime\prime}}{U} \left(  5 P_2 + 2 X_1 - 3 X_2 - X_{3a} - 7 \right) \\
       & + J_2^{\prime\prime} \left(  4 P_2 + P_4 - 8 X_2 - 2 \right)
\end{split}
\end{equation}
and
\begin{equation}
\begin{split}
    B & =  - J_1 P_1  + 16 J_2 P_1  \\
        & - \frac{4 J_1}{U^2} 
        \left[ \left( 6 t^{\prime 2}-  t^\prime t^{\prime\prime} \right) \left( X_1- 1 \right) +  3 t^{\prime\prime 2} \left( P_2 - 2 \right) \right] P_1
\end{split}
\end{equation}
where $J_1 = \frac{4 t^2}{U}$, $J_2 = \frac{4 t^4}{U^3}$, $J_1^\prime = \frac{4 
t^{\prime 2}}{U}$, $J_2^\prime = \frac{4 t^{\prime 4}}{U^3}$, 
$J_1^{\prime\prime} = \frac{4 t^{\prime\prime 2}}{U}$ and $J_2^{\prime\prime} = 
\frac{4 t^{\prime\prime 4}}{U^3}$. When neglecting higher order terms (i.e. 
terms in $J_2^\prime$, $J_2^{\prime\prime}$ and $J_1^\prime J_1^{\prime\prime}$) 
and considering $t^{\prime\prime}=-t^\prime/2$, it is possible to obtain an 
approximated solution for the zone-boundary dispersion 
$E_{ZB}$~\cite{IvashkoPRB2017}:
\begin{equation}
    \frac{E_{ZB}}{12 Z J_2} \approx 1 + \frac{1}{12} \left( 112 - \frac{J_1}{J_2} \right) \left( \frac{t^\prime}{t} \right)^2,
\end{equation}
if:
\begin{equation*}
    \frac{U}{t} \geqslant \sqrt{\frac{28 + 112 \left( \frac{t^\prime}{t} \right)^2}{2 + 3 \left( \frac{t^\prime}{t} \right)^2}}, ~~~ \mathrm{and} ~~~ \left|     \frac{t^\prime}{t} \right| \lesssim  0.686. 
\end{equation*}
Furthermore, it is possible to see~\cite{DelannoyPRB2009}, that with such a 
model, which considers also the cyclic hopping terms, the effective exchange 
interaction can be written as:
\begin{equation}\label{eq:Jeff}
    J_\mathrm{eff}=4\frac{t^2}{U}-64\frac{t^4}{U^3}
\end{equation}
if considering only the first neighbour hopping $t$.\\

{\it DFT and cRPA Calculations:}
We compute the electronic structure of tetragonal bulk La$_2$CuO$_4${} for lattice 
constants and atomic positions corresponding to the experimentally investigated 
thin films (see Table~\ref{tab:tab1}).
For simplicity, 
tetragonal structures were considered with the ratio between copper
to apical oxygen $d_\mathrm{O2}$ (copper to lanthanum $d_\mathrm{La}$) distance 
and the $c$ axis kept constant to the bulk values ${d_\mathrm{O2}}/{c}=0.18(4)$ 
($d_\mathrm{La}/c=0.36(1)$)~\cite{RadaelliPRB94}. We use a full-potential linear muffin-tin orbitals (FPLMTO) implementation~\cite{fplmto} in the local 
density approximation (LDA) and construct maximally localized Wannier functions 
\cite{MarzariRMP2012} for the Cu $d_{x^2-y^2}$ orbital. Hopping elements $t$, 
$t^\prime$ and $t^{\prime\prime}$ are then extracted by fitting a square-lattice 
dispersion to high-symmetry points. 
Next, the static Hubbard $U=U(\omega=0)$ 
is computed using the constrained random-phase approximation (cRPA) 
\cite{AryasetiawanPRB2004} in the Wannier setup \cite{MiyakePRB2008} for 
entangled band-structures \cite{MiyakePRB2009}.
Finally, the effective magnetic exchange interaction $J_\mathrm{eff}$ is determined using the strong-coupling expression Eq.~\ref{eq:Jeff}.

The above procedure is an approximate way to account for the screening of the Coulomb interaction $v$ provided by the electronic degrees of
freedom that are omitted when going to a description in terms of an effective one-band Hubbard (and, ultimately, Heisenberg) model.
In other words, $v$ has to be screened by all particle-hole polarizations
that are not fully contained in the subspace spanned by the $d_{x^2-y^2}$ Wannier functions that define the low-energy model.
In cRPA, this partial polarization is computed within RPA, meaning that bare particle-hole bubble diagrams
(Lindhard function) are summed up to all orders in the interaction.
Constraining the polarization comes with the benefit that it is precisely the left-out
low-energy excitations that display the most correlation effects, potentially leading to important vertex corrections
beyond the RPA. Indeed, solving the many-body model that we are setting up through the hoppings $t$, $t^\prime$, etc.\ and the Hubbard $U$ would require
approaches beyond the RPA.

Let us briefly describe how pressure can modify the partially screened local Coulomb interaction $U$: 
First, pressure-induced changes in hoppings and crystal-fields modify the solid's polarization (dielectric function) and, hence, how efficiently the Coulomb interaction is screened.
This effect is very material specific and can lead to both, an enhancement or a diminishing of $U$~\cite{TomczakPRB2010,SantePRB2017,KimPRB2018}.
Second, the parameters of the Hubbard model are basis-dependent quantities. As a result the matrix element $U$ also depends
on the extent in real-space of the $d_{x^2-y^2}$-derived Wannier basis.
Quite counter-intuitively, a pressure-induced delocalization of Wannier functions
generally leads to increased local interactions~\cite{TomczakPRB2009}.
This trend can be illustrated by looking at the pressure evolution of the matrix element of the bare (unscreened) Coulomb interaction $v=e^2/r$
in the Wannier basis: Indeed, as reported in Table~\ref{tab:tab1}, $v$ increases with shrinking lattice constant.
In our case of tetragonal La$_2$CuO$_4$, both effects (screening and basis localization) promote the same tendency: an increase of the Hubbard $U$ under compression. \\

\textbf{Data availability.}
All experimental data are available upon request to the corresponding authors. The source data underlying Figures.~1-3, Table~\ref{tab:tab1}, Supplementary Figures~1-4 and Supplementary Table~1 are provided as a Source Data file.\\

\textbf{Acknowledgments:}
O.I., M.H. and J.C. acknowledge support by the Swiss National Science Foundation under grant No. BSSGI0$\_155873$ and through the SINERGIA network Mott Physics Beyond the Heisenberg Model. D.E.N., E.P., Y.T. and T.S. acknowledge support by the Swiss National Science Foundation through its Sinergia network Mott Physics Beyond the Heisenberg Model – MPBH (Research Grant CRSII2\_160765/1)  and the NCCR MARVEL (Research Grant 51NF40\_141828).
This work was performed at the ADRESS beamline of the SLS at the Paul Scherrer Institut, Villigen PSI, Switzerland. We thank the ADRESS beamline staff for technical support.
C.A. and M.R.B. are supported by Air Force Office of Scientific Research grant No. FA9550-09-1-0583.
N.E.S. and H.M.R. acknowledge the Swiss National Science foundation under grant No. 200021-169061.
W.W. and N.B.C. were supported by the Danish Center for Synchrotron and Neutron Science (DanScatt). K.M.S. and H.I.W. were supported by the Air Force Office of Scientific Research grant No. FA9550-15-1-0474.\\

\textbf{Author contributions:}
H.I.W., C.A., C.L., M.G., M.R.B. and K.M.S. grew and characterised the La$_2$CuO$_4${} thin films. 
O.I., M.H., W.W., N.B.C., D.E.N., E.P., Y.T., T.S. and J.C. executed the XAS and 
RIXS experiments. O.I., W.W., and N.B.C. performed the RIXS and XAS data 
analysis. N.E.S. and H.M.R. developed the Hubbard model. J.M.T. carried out the DFT and cRPA calculations.  All authors 
contributed to the manuscript. O.I. and M.H. contributed equally to this 
work.\\

\textbf{Competing interests:} The authors declare no competing interests.\\

\vspace{2mm}

\bibliography{RIXS}

\begin{thebibliography}{53}%
\makeatletter
\providecommand \@ifxundefined [1]{%
 \@ifx{#1\undefined}
}%
\providecommand \@ifnum [1]{%
 \ifnum #1\expandafter \@firstoftwo
 \else \expandafter \@secondoftwo
 \fi
}%
\providecommand \@ifx [1]{%
 \ifx #1\expandafter \@firstoftwo
 \else \expandafter \@secondoftwo
 \fi
}%
\providecommand \natexlab [1]{#1}%
\providecommand \enquote  [1]{``#1''}%
\providecommand \bibnamefont  [1]{#1}%
\providecommand \bibfnamefont [1]{#1}%
\providecommand \citenamefont [1]{#1}%
\providecommand \href@noop [0]{\@secondoftwo}%
\providecommand \href [0]{\begingroup \@sanitize@url \@href}%
\providecommand \@href[1]{\@@startlink{#1}\@@href}%
\providecommand \@@href[1]{\endgroup#1\@@endlink}%
\providecommand \@sanitize@url [0]{\catcode `\\12\catcode `\$12\catcode
  `\&12\catcode `\#12\catcode `\^12\catcode `\_12\catcode `\%12\relax}%
\providecommand \@@startlink[1]{}%
\providecommand \@@endlink[0]{}%
\providecommand \url  [0]{\begingroup\@sanitize@url \@url }%
\providecommand \@url [1]{\endgroup\@href {#1}{\urlprefix }}%
\providecommand \urlprefix  [0]{URL }%
\providecommand \Eprint [0]{\href }%
\providecommand \doibase [0]{http://dx.doi.org/}%
\providecommand \selectlanguage [0]{\@gobble}%
\providecommand \bibinfo  [0]{\@secondoftwo}%
\providecommand \bibfield  [0]{\@secondoftwo}%
\providecommand \translation [1]{[#1]}%
\providecommand \BibitemOpen [0]{}%
\providecommand \bibitemStop [0]{}%
\providecommand \bibitemNoStop [0]{.\EOS\space}%
\providecommand \EOS [0]{\spacefactor3000\relax}%
\providecommand \BibitemShut  [1]{\csname bibitem#1\endcsname}%
\let\auto@bib@innerbib\@empty
\bibitem [{\citenamefont {Imada}\ \emph {et~al.}(1998)\citenamefont {Imada},
  \citenamefont {Fujimori},\ and\ \citenamefont {Tokura}}]{ImadaRMP1998}%
  \BibitemOpen
  \bibfield  {author} {\bibinfo {author} {\bibfnamefont {Masatoshi}\
  \bibnamefont {Imada}}, \bibinfo {author} {\bibfnamefont {Atsushi}\
  \bibnamefont {Fujimori}}, \ and\ \bibinfo {author} {\bibfnamefont
  {Yoshinori}\ \bibnamefont {Tokura}},\ }\bibfield  {title} {\enquote {\bibinfo
  {title} {{Metal-insulator transitions}},}\ }\href {\doibase
  10.1103/RevModPhys.70.1039} {\bibfield  {journal} {\bibinfo  {journal} {Rev.
  Mod. Phys.}\ }\textbf {\bibinfo {volume} {70}},\ \bibinfo {pages}
  {1039--1263} (\bibinfo {year} {1998})}\BibitemShut {NoStop}%
\bibitem [{\citenamefont {Wigner}\ and\ \citenamefont
  {Huntington}(1935)}]{Wigner35}%
  \BibitemOpen
  \bibfield  {author} {\bibinfo {author} {\bibfnamefont {E.}~\bibnamefont
  {Wigner}}\ and\ \bibinfo {author} {\bibfnamefont {H.~B.}\ \bibnamefont
  {Huntington}},\ }\bibfield  {title} {\enquote {\bibinfo {title} {{On the
  Possibility of a Metallic Modification of Hydrogen}},}\ }\href {\doibase
  10.1063/1.1749590} {\bibfield  {journal} {\bibinfo  {journal} {J. Chem.
  Phys.}\ }\textbf {\bibinfo {volume} {3}},\ \bibinfo {pages} {764} (\bibinfo
  {year} {1935})}\BibitemShut {NoStop}%
\bibitem [{\citenamefont {Friedemann}\ \emph {et~al.}(2016)\citenamefont
  {Friedemann}, \citenamefont {Chang}, \citenamefont {Gam{\.z}a}, \citenamefont
  {Reiss}, \citenamefont {Chen}, \citenamefont {Alireza}, \citenamefont
  {Coniglio}, \citenamefont {Graf}, \citenamefont {Tozer},\ and\ \citenamefont
  {Grosche}}]{FriedemannSciRep16}%
  \BibitemOpen
  \bibfield  {author} {\bibinfo {author} {\bibfnamefont {S.}~\bibnamefont
  {Friedemann}}, \bibinfo {author} {\bibfnamefont {H.}~\bibnamefont {Chang}},
  \bibinfo {author} {\bibfnamefont {M.~B.}\ \bibnamefont {Gam{\.z}a}}, \bibinfo
  {author} {\bibfnamefont {P.}~\bibnamefont {Reiss}}, \bibinfo {author}
  {\bibfnamefont {X.}~\bibnamefont {Chen}}, \bibinfo {author} {\bibfnamefont
  {P.}~\bibnamefont {Alireza}}, \bibinfo {author} {\bibfnamefont {W.~A.}\
  \bibnamefont {Coniglio}}, \bibinfo {author} {\bibfnamefont {D.}~\bibnamefont
  {Graf}}, \bibinfo {author} {\bibfnamefont {S.}~\bibnamefont {Tozer}}, \ and\
  \bibinfo {author} {\bibfnamefont {F.~M.}\ \bibnamefont {Grosche}},\
  }\bibfield  {title} {\enquote {\bibinfo {title} {{Large Fermi Surface of
  Heavy Electrons at the Border of Mott Insulating State in NiS2}},}\ }\href
  {http://dx.doi.org/10.1038/srep25335} {\bibfield  {journal} {\bibinfo
  {journal} {Sci. Rep.}\ }\textbf {\bibinfo {volume} {6}},\ \bibinfo {pages}
  {25335 EP --} (\bibinfo {year} {2016})}\BibitemShut {NoStop}%
\bibitem [{\citenamefont {Gati}\ \emph {et~al.}(2016)\citenamefont {Gati},
  \citenamefont {Garst}, \citenamefont {Manna}, \citenamefont {Tutsch},
  \citenamefont {Wolf}, \citenamefont {Bartosch}, \citenamefont {Schubert},
  \citenamefont {Sasaki}, \citenamefont {Schlueter},\ and\ \citenamefont
  {Lang}}]{GatieScienceAdv16}%
  \BibitemOpen
  \bibfield  {author} {\bibinfo {author} {\bibfnamefont {Elena}\ \bibnamefont
  {Gati}}, \bibinfo {author} {\bibfnamefont {Markus}\ \bibnamefont {Garst}},
  \bibinfo {author} {\bibfnamefont {Rudra~S.}\ \bibnamefont {Manna}}, \bibinfo
  {author} {\bibfnamefont {Ulrich}\ \bibnamefont {Tutsch}}, \bibinfo {author}
  {\bibfnamefont {Bernd}\ \bibnamefont {Wolf}}, \bibinfo {author}
  {\bibfnamefont {Lorenz}\ \bibnamefont {Bartosch}}, \bibinfo {author}
  {\bibfnamefont {Harald}\ \bibnamefont {Schubert}}, \bibinfo {author}
  {\bibfnamefont {Takahiko}\ \bibnamefont {Sasaki}}, \bibinfo {author}
  {\bibfnamefont {John~A.}\ \bibnamefont {Schlueter}}, \ and\ \bibinfo {author}
  {\bibfnamefont {Michael}\ \bibnamefont {Lang}},\ }\bibfield  {title}
  {\enquote {\bibinfo {title} {{Breakdown of Hooke{\textquoteright}s law of
  elasticity at the Mott critical endpoint in an organic conductor}},}\ }\href
  {\doibase 10.1126/sciadv.1601646} {\bibfield  {journal} {\bibinfo  {journal}
  {Sci. Adv.}\ }\textbf {\bibinfo {volume} {2}} (\bibinfo {year} {2016}),\
  10.1126/sciadv.1601646}\BibitemShut {NoStop}%
\bibitem [{\citenamefont {Lee}\ \emph {et~al.}(2006)\citenamefont {Lee},
  \citenamefont {Nagaosa},\ and\ \citenamefont {Wen}}]{LeeRMP06}%
  \BibitemOpen
  \bibfield  {author} {\bibinfo {author} {\bibfnamefont {Patrick~A.}\
  \bibnamefont {Lee}}, \bibinfo {author} {\bibfnamefont {Naoto}\ \bibnamefont
  {Nagaosa}}, \ and\ \bibinfo {author} {\bibfnamefont {Xiao-Gang}\ \bibnamefont
  {Wen}},\ }\bibfield  {title} {\enquote {\bibinfo {title} {{Doping a Mott
  insulator: Physics of high-temperature superconductivity}},}\ }\href
  {\doibase 10.1103/RevModPhys.78.17} {\bibfield  {journal} {\bibinfo
  {journal} {Rev. Mod. Phys.}\ }\textbf {\bibinfo {volume} {78}},\ \bibinfo
  {pages} {17--85} (\bibinfo {year} {2006})}\BibitemShut {NoStop}%
\bibitem [{\citenamefont {Hardy}\ \emph {et~al.}(2010)\citenamefont {Hardy},
  \citenamefont {Hillier}, \citenamefont {Meingast}, \citenamefont {Colson},
  \citenamefont {Li}, \citenamefont {Bari\ifmmode \check{s}\else
  \v{s}\fi{}i\ifmmode~\acute{c}\else \'{c}\fi{}}, \citenamefont {Yu},
  \citenamefont {Zhao}, \citenamefont {Greven},\ and\ \citenamefont
  {Schilling}}]{HardyPRL2010}%
  \BibitemOpen
  \bibfield  {author} {\bibinfo {author} {\bibfnamefont {F.}~\bibnamefont
  {Hardy}}, \bibinfo {author} {\bibfnamefont {N.~J.}\ \bibnamefont {Hillier}},
  \bibinfo {author} {\bibfnamefont {C.}~\bibnamefont {Meingast}}, \bibinfo
  {author} {\bibfnamefont {D.}~\bibnamefont {Colson}}, \bibinfo {author}
  {\bibfnamefont {Y.}~\bibnamefont {Li}}, \bibinfo {author} {\bibfnamefont
  {N.}~\bibnamefont {Bari\ifmmode \check{s}\else
  \v{s}\fi{}i\ifmmode~\acute{c}\else \'{c}\fi{}}}, \bibinfo {author}
  {\bibfnamefont {G.}~\bibnamefont {Yu}}, \bibinfo {author} {\bibfnamefont
  {X.}~\bibnamefont {Zhao}}, \bibinfo {author} {\bibfnamefont {M.}~\bibnamefont
  {Greven}}, \ and\ \bibinfo {author} {\bibfnamefont {J.~S.}\ \bibnamefont
  {Schilling}},\ }\bibfield  {title} {\enquote {\bibinfo {title} {{Enhancement
  of the Critical Temperature of
  ${\mathrm{HgBa}}_{2}{\mathrm{CuO}}_{4+\ensuremath{\delta}}$ by Applying
  Uniaxial and Hydrostatic Pressure: Implications for a Universal Trend in
  Cuprate Superconductors}},}\ }\href {\doibase 10.1103/PhysRevLett.105.167002}
  {\bibfield  {journal} {\bibinfo  {journal} {Phys. Rev. Lett.}\ }\textbf
  {\bibinfo {volume} {105}},\ \bibinfo {pages} {167002} (\bibinfo {year}
  {2010})}\BibitemShut {NoStop}%
\bibitem [{\citenamefont {Wang}\ \emph {et~al.}(2014)\citenamefont {Wang},
  \citenamefont {Zhang}, \citenamefont {Yan}, \citenamefont {Chen},
  \citenamefont {Struzhkin}, \citenamefont {Tabis}, \citenamefont {Bari\ifmmode
  \check{s}\else \v{s}\fi{}i\ifmmode~\acute{c}\else \'{c}\fi{}}, \citenamefont
  {Chan}, \citenamefont {Dorow}, \citenamefont {Zhao}, \citenamefont {Greven},
  \citenamefont {Mao},\ and\ \citenamefont {Geballe}}]{WangPRB2014}%
  \BibitemOpen
  \bibfield  {author} {\bibinfo {author} {\bibfnamefont {Shibing}\ \bibnamefont
  {Wang}}, \bibinfo {author} {\bibfnamefont {Jianbo}\ \bibnamefont {Zhang}},
  \bibinfo {author} {\bibfnamefont {Jinyuan}\ \bibnamefont {Yan}}, \bibinfo
  {author} {\bibfnamefont {Xiao-Jia}\ \bibnamefont {Chen}}, \bibinfo {author}
  {\bibfnamefont {Viktor}\ \bibnamefont {Struzhkin}}, \bibinfo {author}
  {\bibfnamefont {Wojciech}\ \bibnamefont {Tabis}}, \bibinfo {author}
  {\bibfnamefont {Neven}\ \bibnamefont {Bari\ifmmode \check{s}\else
  \v{s}\fi{}i\ifmmode~\acute{c}\else \'{c}\fi{}}}, \bibinfo {author}
  {\bibfnamefont {Mun~K.}\ \bibnamefont {Chan}}, \bibinfo {author}
  {\bibfnamefont {Chelsey}\ \bibnamefont {Dorow}}, \bibinfo {author}
  {\bibfnamefont {Xudong}\ \bibnamefont {Zhao}}, \bibinfo {author}
  {\bibfnamefont {Martin}\ \bibnamefont {Greven}}, \bibinfo {author}
  {\bibfnamefont {Wendy~L.}\ \bibnamefont {Mao}}, \ and\ \bibinfo {author}
  {\bibfnamefont {Ted}\ \bibnamefont {Geballe}},\ }\bibfield  {title} {\enquote
  {\bibinfo {title} {{Strain derivatives of ${T}_{c}$ in
  HgBa${}_{2}$CuO${}_{4+\ensuremath{\delta}}$: The CuO${}_{2}$ plane alone is
  not enough}},}\ }\href {\doibase 10.1103/PhysRevB.89.024515} {\bibfield
  {journal} {\bibinfo  {journal} {Phys. Rev. B}\ }\textbf {\bibinfo {volume}
  {89}},\ \bibinfo {pages} {024515} (\bibinfo {year} {2014})}\BibitemShut
  {NoStop}%
\bibitem [{\citenamefont {Sala}\ \emph {et~al.}(2011)\citenamefont {Sala},
  \citenamefont {Bisogni}, \citenamefont {Aruta}, \citenamefont {Balestrino},
  \citenamefont {Berger}, \citenamefont {Brookes}, \citenamefont {de~Luca},
  \citenamefont {Castro}, \citenamefont {Grioni}, \citenamefont {Guarise},
  \citenamefont {Medaglia}, \citenamefont {Granozio}, \citenamefont {Minola},
  \citenamefont {Perna}, \citenamefont {Radovic}, \citenamefont {Salluzzo},
  \citenamefont {Schmitt}, \citenamefont {Zhou}, \citenamefont {Braicovich},\
  and\ \citenamefont {Ghiringhelli}}]{MorettiNJP11}%
  \BibitemOpen
  \bibfield  {author} {\bibinfo {author} {\bibfnamefont {M.~Moretti}\
  \bibnamefont {Sala}}, \bibinfo {author} {\bibfnamefont {V.}~\bibnamefont
  {Bisogni}}, \bibinfo {author} {\bibfnamefont {C.}~\bibnamefont {Aruta}},
  \bibinfo {author} {\bibfnamefont {G.}~\bibnamefont {Balestrino}}, \bibinfo
  {author} {\bibfnamefont {H.}~\bibnamefont {Berger}}, \bibinfo {author}
  {\bibfnamefont {N.~B.}\ \bibnamefont {Brookes}}, \bibinfo {author}
  {\bibfnamefont {G.~M.}\ \bibnamefont {de~Luca}}, \bibinfo {author}
  {\bibfnamefont {D.~Di}\ \bibnamefont {Castro}}, \bibinfo {author}
  {\bibfnamefont {M.}~\bibnamefont {Grioni}}, \bibinfo {author} {\bibfnamefont
  {M.}~\bibnamefont {Guarise}}, \bibinfo {author} {\bibfnamefont {P.~G.}\
  \bibnamefont {Medaglia}}, \bibinfo {author} {\bibfnamefont {F.~Miletto}\
  \bibnamefont {Granozio}}, \bibinfo {author} {\bibfnamefont {M.}~\bibnamefont
  {Minola}}, \bibinfo {author} {\bibfnamefont {P.}~\bibnamefont {Perna}},
  \bibinfo {author} {\bibfnamefont {M.}~\bibnamefont {Radovic}}, \bibinfo
  {author} {\bibfnamefont {M.}~\bibnamefont {Salluzzo}}, \bibinfo {author}
  {\bibfnamefont {T.}~\bibnamefont {Schmitt}}, \bibinfo {author} {\bibfnamefont
  {K.~J.}\ \bibnamefont {Zhou}}, \bibinfo {author} {\bibfnamefont
  {L.}~\bibnamefont {Braicovich}}, \ and\ \bibinfo {author} {\bibfnamefont
  {G.}~\bibnamefont {Ghiringhelli}},\ }\bibfield  {title} {\enquote {\bibinfo
  {title} {{Energy and symmetry of dd excitations in undoped layered cuprates
  measured by Cu L 3 resonant inelastic x-ray scattering}},}\ }\href
  {http://stacks.iop.org/1367-2630/13/i=4/a=043026} {\bibfield  {journal}
  {\bibinfo  {journal} {New J. Phys.}\ }\textbf {\bibinfo {volume} {13}},\
  \bibinfo {pages} {043026} (\bibinfo {year} {2011})}\BibitemShut {NoStop}%
\bibitem [{\citenamefont {H\"afliger}\ \emph {et~al.}(2014)\citenamefont
  {H\"afliger}, \citenamefont {Gerber}, \citenamefont {Pramod}, \citenamefont
  {Schnells}, \citenamefont {Piazza}, \citenamefont {Chati}, \citenamefont
  {Pomjakushin}, \citenamefont {Conder}, \citenamefont {Pomjakushina},
  \citenamefont {Le~Dreau}, \citenamefont {Christensen}, \citenamefont
  {Sylju\aa{}sen}, \citenamefont {Normand},\ and\ \citenamefont
  {R\o{}nnow}}]{HafligerPRB2014}%
  \BibitemOpen
  \bibfield  {author} {\bibinfo {author} {\bibfnamefont {P.~S.}\ \bibnamefont
  {H\"afliger}}, \bibinfo {author} {\bibfnamefont {S.}~\bibnamefont {Gerber}},
  \bibinfo {author} {\bibfnamefont {R.}~\bibnamefont {Pramod}}, \bibinfo
  {author} {\bibfnamefont {V.~I.}\ \bibnamefont {Schnells}}, \bibinfo {author}
  {\bibfnamefont {B.~dalla}\ \bibnamefont {Piazza}}, \bibinfo {author}
  {\bibfnamefont {R.}~\bibnamefont {Chati}}, \bibinfo {author} {\bibfnamefont
  {V.}~\bibnamefont {Pomjakushin}}, \bibinfo {author} {\bibfnamefont
  {K.}~\bibnamefont {Conder}}, \bibinfo {author} {\bibfnamefont
  {E.}~\bibnamefont {Pomjakushina}}, \bibinfo {author} {\bibfnamefont
  {L.}~\bibnamefont {Le~Dreau}}, \bibinfo {author} {\bibfnamefont {N.~B.}\
  \bibnamefont {Christensen}}, \bibinfo {author} {\bibfnamefont {O.~F.}\
  \bibnamefont {Sylju\aa{}sen}}, \bibinfo {author} {\bibfnamefont
  {B.}~\bibnamefont {Normand}}, \ and\ \bibinfo {author} {\bibfnamefont
  {H.~M.}\ \bibnamefont {R\o{}nnow}},\ }\bibfield  {title} {\enquote {\bibinfo
  {title} {{Quantum and thermal ionic motion, oxygen isotope effect, and
  superexchange distribution in ${\text{La}}_{2}{\text{CuO}}_{4}$}},}\ }\href
  {\doibase 10.1103/PhysRevB.89.085113} {\bibfield  {journal} {\bibinfo
  {journal} {Phys. Rev. B}\ }\textbf {\bibinfo {volume} {89}},\ \bibinfo
  {pages} {085113} (\bibinfo {year} {2014})}\BibitemShut {NoStop}%
\bibitem [{\citenamefont {Peng}\ \emph {et~al.}(2017)\citenamefont {Peng},
  \citenamefont {Dellea}, \citenamefont {Minola}, \citenamefont {Conni},
  \citenamefont {Amorese}, \citenamefont {Di~Castro}, \citenamefont {De~Luca},
  \citenamefont {Kummer}, \citenamefont {Salluzzo}, \citenamefont {Sun},
  \citenamefont {Zhou}, \citenamefont {Balestrino}, \citenamefont {Le~Tacon},
  \citenamefont {Keimer}, \citenamefont {Braicovich}, \citenamefont {Brookes},\
  and\ \citenamefont {Ghiringhelli}}]{PengNatPhys2016}%
  \BibitemOpen
  \bibfield  {author} {\bibinfo {author} {\bibfnamefont {Y.~Y.}\ \bibnamefont
  {Peng}}, \bibinfo {author} {\bibfnamefont {G.}~\bibnamefont {Dellea}},
  \bibinfo {author} {\bibfnamefont {M.}~\bibnamefont {Minola}}, \bibinfo
  {author} {\bibfnamefont {M.}~\bibnamefont {Conni}}, \bibinfo {author}
  {\bibfnamefont {A.}~\bibnamefont {Amorese}}, \bibinfo {author} {\bibfnamefont
  {D.}~\bibnamefont {Di~Castro}}, \bibinfo {author} {\bibfnamefont {G.~M.}\
  \bibnamefont {De~Luca}}, \bibinfo {author} {\bibfnamefont {K.}~\bibnamefont
  {Kummer}}, \bibinfo {author} {\bibfnamefont {M.}~\bibnamefont {Salluzzo}},
  \bibinfo {author} {\bibfnamefont {X.}~\bibnamefont {Sun}}, \bibinfo {author}
  {\bibfnamefont {X.~J.}\ \bibnamefont {Zhou}}, \bibinfo {author}
  {\bibfnamefont {G.}~\bibnamefont {Balestrino}}, \bibinfo {author}
  {\bibfnamefont {M.}~\bibnamefont {Le~Tacon}}, \bibinfo {author}
  {\bibfnamefont {B.}~\bibnamefont {Keimer}}, \bibinfo {author} {\bibfnamefont
  {L.}~\bibnamefont {Braicovich}}, \bibinfo {author} {\bibfnamefont {N.~B.}\
  \bibnamefont {Brookes}}, \ and\ \bibinfo {author} {\bibfnamefont
  {G.}~\bibnamefont {Ghiringhelli}},\ }\bibfield  {title} {\enquote {\bibinfo
  {title} {{Influence of apical oxygen on the extent of in-plane exchange
  interaction in cuprate superconductors}},}\ }\href
  {http://dx.doi.org/10.1038/nphys4248} {\bibfield  {journal} {\bibinfo
  {journal} {Nat. Phys.}\ }\textbf {\bibinfo {volume} {13}},\ \bibinfo {pages}
  {1201 EP --} (\bibinfo {year} {2017})}\BibitemShut {NoStop}%
\bibitem [{\citenamefont {Ivashko}\ \emph {et~al.}(2017)\citenamefont
  {Ivashko}, \citenamefont {Shaik}, \citenamefont {Lu}, \citenamefont
  {Fatuzzo}, \citenamefont {Dantz}, \citenamefont {Freeman}, \citenamefont
  {McNally}, \citenamefont {Destraz}, \citenamefont {Christensen},
  \citenamefont {Kurosawa}, \citenamefont {Momono}, \citenamefont {Oda},
  \citenamefont {Matt}, \citenamefont {Monney}, \citenamefont {R\o{}nnow},
  \citenamefont {Schmitt},\ and\ \citenamefont {Chang}}]{IvashkoPRB2017}%
  \BibitemOpen
  \bibfield  {author} {\bibinfo {author} {\bibfnamefont {O.}~\bibnamefont
  {Ivashko}}, \bibinfo {author} {\bibfnamefont {N.~E.}\ \bibnamefont {Shaik}},
  \bibinfo {author} {\bibfnamefont {X.}~\bibnamefont {Lu}}, \bibinfo {author}
  {\bibfnamefont {C.~G.}\ \bibnamefont {Fatuzzo}}, \bibinfo {author}
  {\bibfnamefont {M.}~\bibnamefont {Dantz}}, \bibinfo {author} {\bibfnamefont
  {P.~G.}\ \bibnamefont {Freeman}}, \bibinfo {author} {\bibfnamefont {D.~E.}\
  \bibnamefont {McNally}}, \bibinfo {author} {\bibfnamefont {D.}~\bibnamefont
  {Destraz}}, \bibinfo {author} {\bibfnamefont {N.~B.}\ \bibnamefont
  {Christensen}}, \bibinfo {author} {\bibfnamefont {T.}~\bibnamefont
  {Kurosawa}}, \bibinfo {author} {\bibfnamefont {N.}~\bibnamefont {Momono}},
  \bibinfo {author} {\bibfnamefont {M.}~\bibnamefont {Oda}}, \bibinfo {author}
  {\bibfnamefont {C.~E.}\ \bibnamefont {Matt}}, \bibinfo {author}
  {\bibfnamefont {C.}~\bibnamefont {Monney}}, \bibinfo {author} {\bibfnamefont
  {H.~M.}\ \bibnamefont {R\o{}nnow}}, \bibinfo {author} {\bibfnamefont
  {T.}~\bibnamefont {Schmitt}}, \ and\ \bibinfo {author} {\bibfnamefont
  {J.}~\bibnamefont {Chang}},\ }\bibfield  {title} {\enquote {\bibinfo {title}
  {{Damped spin excitations in a doped cuprate superconductor with orbital
  hybridization}},}\ }\href {\doibase 10.1103/PhysRevB.95.214508} {\bibfield
  {journal} {\bibinfo  {journal} {Phys. Rev. B}\ }\textbf {\bibinfo {volume}
  {95}},\ \bibinfo {pages} {214508} (\bibinfo {year} {2017})}\BibitemShut
  {NoStop}%
\bibitem [{\citenamefont {Coldea}\ \emph {et~al.}(2001)\citenamefont {Coldea},
  \citenamefont {Hayden}, \citenamefont {Aeppli}, \citenamefont {Perring},
  \citenamefont {Frost}, \citenamefont {Mason}, \citenamefont {Cheong},\ and\
  \citenamefont {Fisk}}]{ColdeaPRL01}%
  \BibitemOpen
  \bibfield  {author} {\bibinfo {author} {\bibfnamefont {R.}~\bibnamefont
  {Coldea}}, \bibinfo {author} {\bibfnamefont {S.~M.}\ \bibnamefont {Hayden}},
  \bibinfo {author} {\bibfnamefont {G.}~\bibnamefont {Aeppli}}, \bibinfo
  {author} {\bibfnamefont {T.~G.}\ \bibnamefont {Perring}}, \bibinfo {author}
  {\bibfnamefont {C.~D.}\ \bibnamefont {Frost}}, \bibinfo {author}
  {\bibfnamefont {T.~E.}\ \bibnamefont {Mason}}, \bibinfo {author}
  {\bibfnamefont {S.-W.}\ \bibnamefont {Cheong}}, \ and\ \bibinfo {author}
  {\bibfnamefont {Z.}~\bibnamefont {Fisk}},\ }\bibfield  {title} {\enquote
  {\bibinfo {title} {{Spin Waves and Electronic Interactions in
  ${\mathrm{La}}_{2}{\mathrm{CuO}}_{4}$}},}\ }\href {\doibase
  10.1103/PhysRevLett.86.5377} {\bibfield  {journal} {\bibinfo  {journal}
  {Phys. Rev. Lett.}\ }\textbf {\bibinfo {volume} {86}},\ \bibinfo {pages}
  {5377--5380} (\bibinfo {year} {2001})}\BibitemShut {NoStop}%
\bibitem [{\citenamefont {Headings}\ \emph {et~al.}(2010)\citenamefont
  {Headings}, \citenamefont {Hayden}, \citenamefont {Coldea},\ and\
  \citenamefont {Perring}}]{HeadingsPRL10}%
  \BibitemOpen
  \bibfield  {author} {\bibinfo {author} {\bibfnamefont {N.~S.}\ \bibnamefont
  {Headings}}, \bibinfo {author} {\bibfnamefont {S.~M.}\ \bibnamefont
  {Hayden}}, \bibinfo {author} {\bibfnamefont {R.}~\bibnamefont {Coldea}}, \
  and\ \bibinfo {author} {\bibfnamefont {T.~G.}\ \bibnamefont {Perring}},\
  }\bibfield  {title} {\enquote {\bibinfo {title} {{Anomalous High-Energy Spin
  Excitations in the High-${T}_{c}$ Superconductor-Parent Antiferromagnet
  ${\mathrm{La}}_{2}{\mathrm{CuO}}_{4}$}},}\ }\href {\doibase
  10.1103/PhysRevLett.105.247001} {\bibfield  {journal} {\bibinfo  {journal}
  {Phys. Rev. Lett.}\ }\textbf {\bibinfo {volume} {105}},\ \bibinfo {pages}
  {247001} (\bibinfo {year} {2010})}\BibitemShut {NoStop}%
\bibitem [{\citenamefont {Tacon}\ \emph {et~al.}(2011)\citenamefont {Tacon},
  \citenamefont {Ghiringhelli}, \citenamefont {Chaloupka}, \citenamefont
  {Sala}, \citenamefont {Hinkov}, \citenamefont {Haverkort}, \citenamefont
  {Minola}, \citenamefont {Bakr}, \citenamefont {Zhou}, \citenamefont
  {Blanco-Canosa}, \citenamefont {Monney}, \citenamefont {Song}, \citenamefont
  {Sun}, \citenamefont {Lin}, \citenamefont {Luca}, \citenamefont {Salluzzo},
  \citenamefont {Khaliullin}, \citenamefont {Schmitt}, \citenamefont
  {Braicovich},\ and\ \citenamefont {Keimer}}]{TaconNATP11}%
  \BibitemOpen
  \bibfield  {author} {\bibinfo {author} {\bibfnamefont {M.~Le}\ \bibnamefont
  {Tacon}}, \bibinfo {author} {\bibfnamefont {G.}~\bibnamefont {Ghiringhelli}},
  \bibinfo {author} {\bibfnamefont {J.}~\bibnamefont {Chaloupka}}, \bibinfo
  {author} {\bibfnamefont {M.~Moretti}\ \bibnamefont {Sala}}, \bibinfo {author}
  {\bibfnamefont {V.}~\bibnamefont {Hinkov}}, \bibinfo {author} {\bibfnamefont
  {M.~W.}\ \bibnamefont {Haverkort}}, \bibinfo {author} {\bibfnamefont
  {M.}~\bibnamefont {Minola}}, \bibinfo {author} {\bibfnamefont
  {M.}~\bibnamefont {Bakr}}, \bibinfo {author} {\bibfnamefont {K.~J.}\
  \bibnamefont {Zhou}}, \bibinfo {author} {\bibfnamefont {S.}~\bibnamefont
  {Blanco-Canosa}}, \bibinfo {author} {\bibfnamefont {C.}~\bibnamefont
  {Monney}}, \bibinfo {author} {\bibfnamefont {Y.~T.}\ \bibnamefont {Song}},
  \bibinfo {author} {\bibfnamefont {G.~L.}\ \bibnamefont {Sun}}, \bibinfo
  {author} {\bibfnamefont {C.~T.}\ \bibnamefont {Lin}}, \bibinfo {author}
  {\bibfnamefont {G.~M.~De}\ \bibnamefont {Luca}}, \bibinfo {author}
  {\bibfnamefont {M.}~\bibnamefont {Salluzzo}}, \bibinfo {author}
  {\bibfnamefont {G.}~\bibnamefont {Khaliullin}}, \bibinfo {author}
  {\bibfnamefont {T.}~\bibnamefont {Schmitt}}, \bibinfo {author} {\bibfnamefont
  {L.}~\bibnamefont {Braicovich}}, \ and\ \bibinfo {author} {\bibfnamefont
  {B.}~\bibnamefont {Keimer}},\ }\bibfield  {title} {\enquote {\bibinfo {title}
  {{Intense paramagnon excitations in a large family of high-temperature
  superconductors}},}\ }\href {\doibase 10.1038/nphys2041} {\bibfield
  {journal} {\bibinfo  {journal} {Nat. Phys.}\ }\textbf {\bibinfo {volume}
  {7}},\ \bibinfo {pages} {725--730} (\bibinfo {year} {2011})}\BibitemShut
  {NoStop}%
\bibitem [{\citenamefont {Monney}\ \emph {et~al.}(2016)\citenamefont {Monney},
  \citenamefont {Schmitt}, \citenamefont {Matt}, \citenamefont {Mesot},
  \citenamefont {Strocov}, \citenamefont {Lipscombe}, \citenamefont {Hayden},\
  and\ \citenamefont {Chang}}]{MonneyPRB16}%
  \BibitemOpen
  \bibfield  {author} {\bibinfo {author} {\bibfnamefont {C.}~\bibnamefont
  {Monney}}, \bibinfo {author} {\bibfnamefont {T.}~\bibnamefont {Schmitt}},
  \bibinfo {author} {\bibfnamefont {C.~E.}\ \bibnamefont {Matt}}, \bibinfo
  {author} {\bibfnamefont {J.}~\bibnamefont {Mesot}}, \bibinfo {author}
  {\bibfnamefont {V.~N.}\ \bibnamefont {Strocov}}, \bibinfo {author}
  {\bibfnamefont {O.~J.}\ \bibnamefont {Lipscombe}}, \bibinfo {author}
  {\bibfnamefont {S.~M.}\ \bibnamefont {Hayden}}, \ and\ \bibinfo {author}
  {\bibfnamefont {J.}~\bibnamefont {Chang}},\ }\bibfield  {title} {\enquote
  {\bibinfo {title} {{Resonant inelastic x-ray scattering study of the spin and
  charge excitations in the overdoped superconductor
  ${\mathrm{La}}_{1.77}{\mathrm{Sr}}_{0.23}{\mathrm{CuO}}_{4}$}},}\ }\href
  {\doibase 10.1103/PhysRevB.93.075103} {\bibfield  {journal} {\bibinfo
  {journal} {Phys. Rev. B}\ }\textbf {\bibinfo {volume} {93}},\ \bibinfo
  {pages} {075103} (\bibinfo {year} {2016})}\BibitemShut {NoStop}%
\bibitem [{\citenamefont {Lamsal}\ and\ \citenamefont
  {Montfrooij}(2016)}]{LamsalPRB16}%
  \BibitemOpen
  \bibfield  {author} {\bibinfo {author} {\bibfnamefont {Jagat}\ \bibnamefont
  {Lamsal}}\ and\ \bibinfo {author} {\bibfnamefont {Wouter}\ \bibnamefont
  {Montfrooij}},\ }\bibfield  {title} {\enquote {\bibinfo {title} {{Extracting
  paramagnon excitations from resonant inelastic x-ray scattering
  experiments}},}\ }\href {\doibase 10.1103/PhysRevB.93.214513} {\bibfield
  {journal} {\bibinfo  {journal} {Phys. Rev. B}\ }\textbf {\bibinfo {volume}
  {93}},\ \bibinfo {pages} {214513} (\bibinfo {year} {2016})}\BibitemShut
  {NoStop}%
\bibitem [{\citenamefont {{M. P. M. Dean}}\ \emph {et~al.}(2013)\citenamefont
  {{M. P. M. Dean}}, \citenamefont {{G. Dellea}}, \citenamefont {{R. S.
  Springell}}, \citenamefont {{F. Yakhou-Harris}}, \citenamefont {{K. Kummer}},
  \citenamefont {{N. B. Brookes}}, \citenamefont {{X. Liu}}, \citenamefont
  {{Y-J. Sun}}, \citenamefont {{J. Strle}}, \citenamefont {{T. Schmitt}},
  \citenamefont {{L. Braicovich}}, \citenamefont {{G. Ghiringhelli}},
  \citenamefont {{I. Bo\v{z}ovi{\'c}}},\ and\ \citenamefont {{J. P.
  Hill}}}]{DeanNATM13}%
  \BibitemOpen
  \bibfield  {author} {\bibinfo {author} {\bibnamefont {{M. P. M. Dean}}},
  \bibinfo {author} {\bibnamefont {{G. Dellea}}}, \bibinfo {author}
  {\bibnamefont {{R. S. Springell}}}, \bibinfo {author} {\bibnamefont {{F.
  Yakhou-Harris}}}, \bibinfo {author} {\bibnamefont {{K. Kummer}}}, \bibinfo
  {author} {\bibnamefont {{N. B. Brookes}}}, \bibinfo {author} {\bibnamefont
  {{X. Liu}}}, \bibinfo {author} {\bibnamefont {{Y-J. Sun}}}, \bibinfo {author}
  {\bibnamefont {{J. Strle}}}, \bibinfo {author} {\bibnamefont {{T. Schmitt}}},
  \bibinfo {author} {\bibnamefont {{L. Braicovich}}}, \bibinfo {author}
  {\bibnamefont {{G. Ghiringhelli}}}, \bibinfo {author} {\bibnamefont {{I.
  Bo\v{z}ovi{\'c}}}}, \ and\ \bibinfo {author} {\bibnamefont {{J. P. Hill}}},\
  }\bibfield  {title} {\enquote {\bibinfo {title} {{Persistence of magnetic
  excitations in La$_{2\ensuremath{-}x}$Sr$_x$CuO$_4$ from the undoped
  insulator to the heavily overdoped non-superconducting metal}},}\ }\href
  {\doibase 10.1038/nmat3723} {\bibfield  {journal} {\bibinfo  {journal} {Nat.
  Mater.}\ }\textbf {\bibinfo {volume} {12}},\ \bibinfo {pages} {1019--1023}
  (\bibinfo {year} {2013})}\BibitemShut {NoStop}%
\bibitem [{\citenamefont {Nomura}(2015)}]{NomuraJPSJ2015}%
  \BibitemOpen
  \bibfield  {author} {\bibinfo {author} {\bibfnamefont {Takuji}\ \bibnamefont
  {Nomura}},\ }\bibfield  {title} {\enquote {\bibinfo {title} {{Theoretical
  Study of L-Edge Resonant Inelastic X-ray Scattering in La2CuO4 on the Basis
  of Detailed Electronic Band Structure}},}\ }\href {\doibase
  10.7566/JPSJ.84.094704} {\bibfield  {journal} {\bibinfo  {journal} {Journal
  of the Physical Society of Japan}\ }\textbf {\bibinfo {volume} {84}},\
  \bibinfo {pages} {094704} (\bibinfo {year} {2015})},\ \Eprint
  {http://arxiv.org/abs/https://doi.org/10.7566/JPSJ.84.094704}
  {https://doi.org/10.7566/JPSJ.84.094704} \BibitemShut {NoStop}%
\bibitem [{\citenamefont {{C. J. Jia}}\ \emph {et~al.}(2014)\citenamefont {{C.
  J. Jia}}, \citenamefont {{E. A. Nowadnick}}, \citenamefont {{K. Wohlfeld}},
  \citenamefont {{Y. F. Kung}}, \citenamefont {{C.-C. Chen}}, \citenamefont
  {{S. Johnston}}, \citenamefont {{T. Tohyama}}, \citenamefont {{B. Moritz}},\
  and\ \citenamefont {{T. P. Devereaux}}}]{JiaNATC2014}%
  \BibitemOpen
  \bibfield  {author} {\bibinfo {author} {\bibnamefont {{C. J. Jia}}}, \bibinfo
  {author} {\bibnamefont {{E. A. Nowadnick}}}, \bibinfo {author} {\bibnamefont
  {{K. Wohlfeld}}}, \bibinfo {author} {\bibnamefont {{Y. F. Kung}}}, \bibinfo
  {author} {\bibnamefont {{C.-C. Chen}}}, \bibinfo {author} {\bibnamefont {{S.
  Johnston}}}, \bibinfo {author} {\bibnamefont {{T. Tohyama}}}, \bibinfo
  {author} {\bibnamefont {{B. Moritz}}}, \ and\ \bibinfo {author} {\bibnamefont
  {{T. P. Devereaux}}},\ }\bibfield  {title} {\enquote {\bibinfo {title}
  {{Persistent spin excitations in doped antiferromagnets revealed by resonant
  inelastic light scattering}},}\ }\href {\doibase 10.1038/ncomms4314}
  {\bibfield  {journal} {\bibinfo  {journal} {Nat. Commun.}\ }\textbf {\bibinfo
  {volume} {5}},\ \bibinfo {pages} {3314} (\bibinfo {year} {2014})}\BibitemShut
  {NoStop}%
\bibitem [{\citenamefont {Delannoy}\ \emph {et~al.}(2009)\citenamefont
  {Delannoy}, \citenamefont {Gingras}, \citenamefont {Holdsworth},\ and\
  \citenamefont {Tremblay}}]{DelannoyPRB2009}%
  \BibitemOpen
  \bibfield  {author} {\bibinfo {author} {\bibfnamefont {J.-Y.~P.}\
  \bibnamefont {Delannoy}}, \bibinfo {author} {\bibfnamefont {M.~J.~P.}\
  \bibnamefont {Gingras}}, \bibinfo {author} {\bibfnamefont {P.~C.~W.}\
  \bibnamefont {Holdsworth}}, \ and\ \bibinfo {author} {\bibfnamefont
  {A.-M.~S.}\ \bibnamefont {Tremblay}},\ }\bibfield  {title} {\enquote
  {\bibinfo {title} {{Low-energy theory of the
  $t\ensuremath{-}{t}^{\ensuremath{'}}\ensuremath{-}{t}^{\ensuremath{''}}\ensuremath{-}U$
  Hubbard model at half-filling: Interaction strengths in cuprate
  superconductors and an effective spin-only description of
  ${\text{La}}_{2}{\text{CuO}}_{4}$}},}\ }\href {\doibase
  10.1103/PhysRevB.79.235130} {\bibfield  {journal} {\bibinfo  {journal} {Phys.
  Rev. B}\ }\textbf {\bibinfo {volume} {79}},\ \bibinfo {pages} {235130}
  (\bibinfo {year} {2009})}\BibitemShut {NoStop}%
\bibitem [{\citenamefont {{Dalla Piazza}}\ \emph {et~al.}(2012)\citenamefont
  {{Dalla Piazza}}, \citenamefont {Mourigal}, \citenamefont {Guarise},
  \citenamefont {Berger}, \citenamefont {Schmitt}, \citenamefont {Zhou},
  \citenamefont {Grioni},\ and\ \citenamefont {R{\o}nnow}}]{DallaPiazzaPRB12}%
  \BibitemOpen
  \bibfield  {author} {\bibinfo {author} {\bibfnamefont {B.}~\bibnamefont
  {{Dalla Piazza}}}, \bibinfo {author} {\bibfnamefont {M.}~\bibnamefont
  {Mourigal}}, \bibinfo {author} {\bibfnamefont {M.}~\bibnamefont {Guarise}},
  \bibinfo {author} {\bibfnamefont {H.}~\bibnamefont {Berger}}, \bibinfo
  {author} {\bibfnamefont {T.}~\bibnamefont {Schmitt}}, \bibinfo {author}
  {\bibfnamefont {K.~J.}\ \bibnamefont {Zhou}}, \bibinfo {author}
  {\bibfnamefont {M.}~\bibnamefont {Grioni}}, \ and\ \bibinfo {author}
  {\bibfnamefont {H.~M.}\ \bibnamefont {R{\o}nnow}},\ }\bibfield  {title}
  {\enquote {\bibinfo {title} {{Unified one-band Hubbard model for magnetic and
  electronic spectra of the parent compounds of cuprate superconductors}},}\
  }\href {\doibase 10.1103/PhysRevB.85.100508} {\bibfield  {journal} {\bibinfo
  {journal} {Phys. Rev. B}\ }\textbf {\bibinfo {volume} {85}},\ \bibinfo
  {pages} {100508} (\bibinfo {year} {2012})}\BibitemShut {NoStop}%
\bibitem [{\citenamefont {Abrecht}\ \emph {et~al.}(2003)\citenamefont
  {Abrecht}, \citenamefont {Ariosa}, \citenamefont {Cloetta}, \citenamefont
  {Mitrovic}, \citenamefont {Onellion}, \citenamefont {Xi}, \citenamefont
  {Margaritondo},\ and\ \citenamefont {Pavuna}}]{AbrechtPRL03}%
  \BibitemOpen
  \bibfield  {author} {\bibinfo {author} {\bibfnamefont {M.}~\bibnamefont
  {Abrecht}}, \bibinfo {author} {\bibfnamefont {D.}~\bibnamefont {Ariosa}},
  \bibinfo {author} {\bibfnamefont {D.}~\bibnamefont {Cloetta}}, \bibinfo
  {author} {\bibfnamefont {S.}~\bibnamefont {Mitrovic}}, \bibinfo {author}
  {\bibfnamefont {M.}~\bibnamefont {Onellion}}, \bibinfo {author}
  {\bibfnamefont {X.~X.}\ \bibnamefont {Xi}}, \bibinfo {author} {\bibfnamefont
  {G.}~\bibnamefont {Margaritondo}}, \ and\ \bibinfo {author} {\bibfnamefont
  {D.}~\bibnamefont {Pavuna}},\ }\bibfield  {title} {\enquote {\bibinfo {title}
  {{Strain and High Temperature Superconductivity: Unexpected Results from
  Direct Electronic Structure Measurements in Thin Films}},}\ }\href {\doibase
  10.1103/PhysRevLett.91.057002} {\bibfield  {journal} {\bibinfo  {journal}
  {Phys. Rev. Lett.}\ }\textbf {\bibinfo {volume} {91}},\ \bibinfo {pages}
  {057002} (\bibinfo {year} {2003})}\BibitemShut {NoStop}%
\bibitem [{\citenamefont {Tomczak}\ \emph {et~al.}(2009)\citenamefont
  {Tomczak}, \citenamefont {Miyake}, \citenamefont {Sakuma},\ and\
  \citenamefont {Aryasetiawan}}]{TomczakPRB2009}%
  \BibitemOpen
  \bibfield  {author} {\bibinfo {author} {\bibfnamefont {Jan~M.}\ \bibnamefont
  {Tomczak}}, \bibinfo {author} {\bibfnamefont {T.}~\bibnamefont {Miyake}},
  \bibinfo {author} {\bibfnamefont {R.}~\bibnamefont {Sakuma}}, \ and\ \bibinfo
  {author} {\bibfnamefont {F.}~\bibnamefont {Aryasetiawan}},\ }\bibfield
  {title} {\enquote {\bibinfo {title} {{Effective Coulomb interactions in
  solids under pressure}},}\ }\href {\doibase 10.1103/PhysRevB.79.235133}
  {\bibfield  {journal} {\bibinfo  {journal} {Phys. Rev. B}\ }\textbf {\bibinfo
  {volume} {79}},\ \bibinfo {pages} {235133} (\bibinfo {year}
  {2009})}\BibitemShut {NoStop}%
\bibitem [{\citenamefont {Tomczak}\ \emph {et~al.}(2010)\citenamefont
  {Tomczak}, \citenamefont {Miyake},\ and\ \citenamefont
  {Aryasetiawan}}]{TomczakPRB2010}%
  \BibitemOpen
  \bibfield  {author} {\bibinfo {author} {\bibfnamefont {Jan~M.}\ \bibnamefont
  {Tomczak}}, \bibinfo {author} {\bibfnamefont {T.}~\bibnamefont {Miyake}}, \
  and\ \bibinfo {author} {\bibfnamefont {F.}~\bibnamefont {Aryasetiawan}},\
  }\bibfield  {title} {\enquote {\bibinfo {title} {{Realistic many-body models
  for manganese monoxide under pressure}},}\ }\href {\doibase
  10.1103/PhysRevB.81.115116} {\bibfield  {journal} {\bibinfo  {journal} {Phys.
  Rev. B}\ }\textbf {\bibinfo {volume} {81}},\ \bibinfo {pages} {115116}
  (\bibinfo {year} {2010})}\BibitemShut {NoStop}%
\bibitem [{\citenamefont {Di~Sante}\ \emph {et~al.}(2017)\citenamefont
  {Di~Sante}, \citenamefont {Hausoel}, \citenamefont {Barone}, \citenamefont
  {Tomczak}, \citenamefont {Sangiovanni},\ and\ \citenamefont
  {Thomale}}]{SantePRB2017}%
  \BibitemOpen
  \bibfield  {author} {\bibinfo {author} {\bibfnamefont {Domenico}\
  \bibnamefont {Di~Sante}}, \bibinfo {author} {\bibfnamefont {Andreas}\
  \bibnamefont {Hausoel}}, \bibinfo {author} {\bibfnamefont {Paolo}\
  \bibnamefont {Barone}}, \bibinfo {author} {\bibfnamefont {Jan~M.}\
  \bibnamefont {Tomczak}}, \bibinfo {author} {\bibfnamefont {Giorgio}\
  \bibnamefont {Sangiovanni}}, \ and\ \bibinfo {author} {\bibfnamefont {Ronny}\
  \bibnamefont {Thomale}},\ }\bibfield  {title} {\enquote {\bibinfo {title}
  {{Realizing double Dirac particles in the presence of electronic
  interactions}},}\ }\href {\doibase 10.1103/PhysRevB.96.121106} {\bibfield
  {journal} {\bibinfo  {journal} {Phys. Rev. B}\ }\textbf {\bibinfo {volume}
  {96}},\ \bibinfo {pages} {121106} (\bibinfo {year} {2017})}\BibitemShut
  {NoStop}%
\bibitem [{\citenamefont {Kim}\ \emph {et~al.}(2018)\citenamefont {Kim},
  \citenamefont {Liu}, \citenamefont {Tomczak},\ and\ \citenamefont
  {Franchini}}]{KimPRB2018}%
  \BibitemOpen
  \bibfield  {author} {\bibinfo {author} {\bibfnamefont {Bongjae}\ \bibnamefont
  {Kim}}, \bibinfo {author} {\bibfnamefont {Peitao}\ \bibnamefont {Liu}},
  \bibinfo {author} {\bibfnamefont {Jan~M.}\ \bibnamefont {Tomczak}}, \ and\
  \bibinfo {author} {\bibfnamefont {Cesare}\ \bibnamefont {Franchini}},\
  }\bibfield  {title} {\enquote {\bibinfo {title} {Strain-induced tuning of the
  electronic coulomb interaction in $3d$ transition metal oxide perovskites},}\
  }\href {\doibase 10.1103/PhysRevB.98.075130} {\bibfield  {journal} {\bibinfo
  {journal} {Phys. Rev. B}\ }\textbf {\bibinfo {volume} {98}},\ \bibinfo
  {pages} {075130} (\bibinfo {year} {2018})}\BibitemShut {NoStop}%
\bibitem [{\citenamefont {Werner}\ \emph {et~al.}(2015)\citenamefont {Werner},
  \citenamefont {Sakuma}, \citenamefont {Nilsson},\ and\ \citenamefont
  {Aryasetiawan}}]{WernerPRB2015}%
  \BibitemOpen
  \bibfield  {author} {\bibinfo {author} {\bibfnamefont {Philipp}\ \bibnamefont
  {Werner}}, \bibinfo {author} {\bibfnamefont {Rei}\ \bibnamefont {Sakuma}},
  \bibinfo {author} {\bibfnamefont {Fredrik}\ \bibnamefont {Nilsson}}, \ and\
  \bibinfo {author} {\bibfnamefont {Ferdi}\ \bibnamefont {Aryasetiawan}},\
  }\bibfield  {title} {\enquote {\bibinfo {title} {{Dynamical screening in
  ${\text{La}}_{2}{\text{CuO}}_{4}$}},}\ }\href {\doibase
  10.1103/PhysRevB.91.125142} {\bibfield  {journal} {\bibinfo  {journal} {Phys.
  Rev. B}\ }\textbf {\bibinfo {volume} {91}},\ \bibinfo {pages} {125142}
  (\bibinfo {year} {2015})}\BibitemShut {NoStop}%
\bibitem [{\citenamefont {Yoshida}\ \emph {et~al.}(2006)\citenamefont
  {Yoshida}, \citenamefont {Zhou}, \citenamefont {Tanaka}, \citenamefont
  {Yang}, \citenamefont {Hussain}, \citenamefont {Shen}, \citenamefont
  {Fujimori}, \citenamefont {Sahrakorpi}, \citenamefont {Lindroos},
  \citenamefont {Markiewicz}, \citenamefont {Bansil}, \citenamefont {Komiya},
  \citenamefont {Ando}, \citenamefont {Eisaki}, \citenamefont {Kakeshita},\
  and\ \citenamefont {Uchida}}]{YoshidaPRB06}%
  \BibitemOpen
  \bibfield  {author} {\bibinfo {author} {\bibfnamefont {T.}~\bibnamefont
  {Yoshida}}, \bibinfo {author} {\bibfnamefont {X.~J.}\ \bibnamefont {Zhou}},
  \bibinfo {author} {\bibfnamefont {K.}~\bibnamefont {Tanaka}}, \bibinfo
  {author} {\bibfnamefont {W.~L.}\ \bibnamefont {Yang}}, \bibinfo {author}
  {\bibfnamefont {Z.}~\bibnamefont {Hussain}}, \bibinfo {author} {\bibfnamefont
  {Z.-X.}\ \bibnamefont {Shen}}, \bibinfo {author} {\bibfnamefont
  {A.}~\bibnamefont {Fujimori}}, \bibinfo {author} {\bibfnamefont
  {S.}~\bibnamefont {Sahrakorpi}}, \bibinfo {author} {\bibfnamefont
  {M.}~\bibnamefont {Lindroos}}, \bibinfo {author} {\bibfnamefont {R.~S.}\
  \bibnamefont {Markiewicz}}, \bibinfo {author} {\bibfnamefont
  {A.}~\bibnamefont {Bansil}}, \bibinfo {author} {\bibfnamefont {Seiki}\
  \bibnamefont {Komiya}}, \bibinfo {author} {\bibfnamefont {Yoichi}\
  \bibnamefont {Ando}}, \bibinfo {author} {\bibfnamefont {H.}~\bibnamefont
  {Eisaki}}, \bibinfo {author} {\bibfnamefont {T.}~\bibnamefont {Kakeshita}}, \
  and\ \bibinfo {author} {\bibfnamefont {S.}~\bibnamefont {Uchida}},\
  }\bibfield  {title} {\enquote {\bibinfo {title} {{Systematic doping evolution
  of the underlying Fermi surface of
  ${\mathrm{La}}_{2\ensuremath{-}x}{\mathrm{Sr}}_{x}\mathrm{Cu}{\mathrm{O}}_{4}$}},}\
  }\href {\doibase 10.1103/PhysRevB.74.224510} {\bibfield  {journal} {\bibinfo
  {journal} {Phys. Rev. B}\ }\textbf {\bibinfo {volume} {74}},\ \bibinfo
  {pages} {224510} (\bibinfo {year} {2006})}\BibitemShut {NoStop}%
\bibitem [{\citenamefont {Sakakibara}\ \emph {et~al.}(2010)\citenamefont
  {Sakakibara}, \citenamefont {Usui}, \citenamefont {Kuroki}, \citenamefont
  {Arita},\ and\ \citenamefont {Aoki}}]{SakakibaraPRL10}%
  \BibitemOpen
  \bibfield  {author} {\bibinfo {author} {\bibfnamefont {Hirofumi}\
  \bibnamefont {Sakakibara}}, \bibinfo {author} {\bibfnamefont {Hidetomo}\
  \bibnamefont {Usui}}, \bibinfo {author} {\bibfnamefont {Kazuhiko}\
  \bibnamefont {Kuroki}}, \bibinfo {author} {\bibfnamefont {Ryotaro}\
  \bibnamefont {Arita}}, \ and\ \bibinfo {author} {\bibfnamefont {Hideo}\
  \bibnamefont {Aoki}},\ }\bibfield  {title} {\enquote {\bibinfo {title}
  {{Two-Orbital Model Explains the Higher Transition Temperature of the
  Single-Layer Hg-Cuprate Superconductor Compared to That of the La-Cuprate
  Superconductor}},}\ }\href {\doibase 10.1103/PhysRevLett.105.057003}
  {\bibfield  {journal} {\bibinfo  {journal} {Phys. Rev. Lett.}\ }\textbf
  {\bibinfo {volume} {105}},\ \bibinfo {pages} {057003} (\bibinfo {year}
  {2010})}\BibitemShut {NoStop}%
\bibitem [{\citenamefont {Sakakibara}\ \emph {et~al.}(2012)\citenamefont
  {Sakakibara}, \citenamefont {Usui}, \citenamefont {Kuroki}, \citenamefont
  {Arita},\ and\ \citenamefont {Aoki}}]{SakakibaraPRB12}%
  \BibitemOpen
  \bibfield  {author} {\bibinfo {author} {\bibfnamefont {Hirofumi}\
  \bibnamefont {Sakakibara}}, \bibinfo {author} {\bibfnamefont {Hidetomo}\
  \bibnamefont {Usui}}, \bibinfo {author} {\bibfnamefont {Kazuhiko}\
  \bibnamefont {Kuroki}}, \bibinfo {author} {\bibfnamefont {Ryotaro}\
  \bibnamefont {Arita}}, \ and\ \bibinfo {author} {\bibfnamefont {Hideo}\
  \bibnamefont {Aoki}},\ }\bibfield  {title} {\enquote {\bibinfo {title}
  {{Origin of the material dependence of ${T}_{c}$ in the single-layered
  cuprates}},}\ }\href {\doibase 10.1103/PhysRevB.85.064501} {\bibfield
  {journal} {\bibinfo  {journal} {Phys. Rev. B}\ }\textbf {\bibinfo {volume}
  {85}},\ \bibinfo {pages} {064501} (\bibinfo {year} {2012})}\BibitemShut
  {NoStop}%
\bibitem [{\citenamefont {Plat{\'e}}\ \emph {et~al.}(2005)\citenamefont
  {Plat{\'e}}, \citenamefont {Mottershead}, \citenamefont {Elfimov},
  \citenamefont {Peets}, \citenamefont {Liang}, \citenamefont {Bonn},
  \citenamefont {Hardy}, \citenamefont {Chiuzbaian}, \citenamefont {Falub},
  \citenamefont {Shi}, \citenamefont {Patthey},\ and\ \citenamefont
  {Damascelli}}]{PlatePRL05}%
  \BibitemOpen
  \bibfield  {author} {\bibinfo {author} {\bibfnamefont {M.}~\bibnamefont
  {Plat{\'e}}}, \bibinfo {author} {\bibfnamefont {J.~D.~F.}\ \bibnamefont
  {Mottershead}}, \bibinfo {author} {\bibfnamefont {I.~S.}\ \bibnamefont
  {Elfimov}}, \bibinfo {author} {\bibfnamefont {D.~C.}\ \bibnamefont {Peets}},
  \bibinfo {author} {\bibfnamefont {Ruixing}\ \bibnamefont {Liang}}, \bibinfo
  {author} {\bibfnamefont {D.~A.}\ \bibnamefont {Bonn}}, \bibinfo {author}
  {\bibfnamefont {W.~N.}\ \bibnamefont {Hardy}}, \bibinfo {author}
  {\bibfnamefont {S.}~\bibnamefont {Chiuzbaian}}, \bibinfo {author}
  {\bibfnamefont {M.}~\bibnamefont {Falub}}, \bibinfo {author} {\bibfnamefont
  {M.}~\bibnamefont {Shi}}, \bibinfo {author} {\bibfnamefont {L.}~\bibnamefont
  {Patthey}}, \ and\ \bibinfo {author} {\bibfnamefont {A.}~\bibnamefont
  {Damascelli}},\ }\bibfield  {title} {\enquote {\bibinfo {title} {{Fermi
  Surface and Quasiparticle Excitations of Overdoped
  ${\mathrm{Tl}}_{2}{\mathrm{Ba}}_{2}{\mathrm{CuO}}_{6+\ensuremath{\delta}}$}},}\
  }\href {\doibase 10.1103/PhysRevLett.95.077001} {\bibfield  {journal}
  {\bibinfo  {journal} {Phys. Rev. Lett.}\ }\textbf {\bibinfo {volume} {95}},\
  \bibinfo {pages} {077001} (\bibinfo {year} {2005})}\BibitemShut {NoStop}%
\bibitem [{\citenamefont {{Matt C. E.}}\ \emph {et~al.}(2018)\citenamefont
  {{Matt C. E.}}, \citenamefont {{Sutter D.}}, \citenamefont {{Cook A. M.}},
  \citenamefont {{Sassa Y.}}, \citenamefont {{Månsson M.}}, \citenamefont
  {{Tjernberg O.}}, \citenamefont {{Das L.}}, \citenamefont {{Horio M.}},
  \citenamefont {{Destraz D.}}, \citenamefont {{Fatuzzo C. G.}}, \citenamefont
  {{Hauser K.}}, \citenamefont {{Shi M.}}, \citenamefont {{Kobayashi M.}},
  \citenamefont {{Strocov V. N.}}, \citenamefont {{Schmitt T.}}, \citenamefont
  {{Dudin P.}}, \citenamefont {{Hoesch M.}}, \citenamefont {{Pyon S.}},
  \citenamefont {{Takayama T.}}, \citenamefont {{Takagi H.}}, \citenamefont
  {{Lipscombe O. J.}}, \citenamefont {{Hayden S. M.}}, \citenamefont {{Kurosawa
  T.}}, \citenamefont {{Momono N.}}, \citenamefont {{Oda M.}}, \citenamefont
  {{Neupert T.}},\ and\ \citenamefont {{Chang J.}}}]{MattNatComm2018}%
  \BibitemOpen
  \bibfield  {author} {\bibinfo {author} {\bibnamefont {{Matt C. E.}}},
  \bibinfo {author} {\bibnamefont {{Sutter D.}}}, \bibinfo {author}
  {\bibnamefont {{Cook A. M.}}}, \bibinfo {author} {\bibnamefont {{Sassa Y.}}},
  \bibinfo {author} {\bibnamefont {{Månsson M.}}}, \bibinfo {author}
  {\bibnamefont {{Tjernberg O.}}}, \bibinfo {author} {\bibnamefont {{Das L.}}},
  \bibinfo {author} {\bibnamefont {{Horio M.}}}, \bibinfo {author}
  {\bibnamefont {{Destraz D.}}}, \bibinfo {author} {\bibnamefont {{Fatuzzo C.
  G.}}}, \bibinfo {author} {\bibnamefont {{Hauser K.}}}, \bibinfo {author}
  {\bibnamefont {{Shi M.}}}, \bibinfo {author} {\bibnamefont {{Kobayashi M.}}},
  \bibinfo {author} {\bibnamefont {{Strocov V. N.}}}, \bibinfo {author}
  {\bibnamefont {{Schmitt T.}}}, \bibinfo {author} {\bibnamefont {{Dudin P.}}},
  \bibinfo {author} {\bibnamefont {{Hoesch M.}}}, \bibinfo {author}
  {\bibnamefont {{Pyon S.}}}, \bibinfo {author} {\bibnamefont {{Takayama T.}}},
  \bibinfo {author} {\bibnamefont {{Takagi H.}}}, \bibinfo {author}
  {\bibnamefont {{Lipscombe O. J.}}}, \bibinfo {author} {\bibnamefont {{Hayden
  S. M.}}}, \bibinfo {author} {\bibnamefont {{Kurosawa T.}}}, \bibinfo {author}
  {\bibnamefont {{Momono N.}}}, \bibinfo {author} {\bibnamefont {{Oda M.}}},
  \bibinfo {author} {\bibnamefont {{Neupert T.}}}, \ and\ \bibinfo {author}
  {\bibnamefont {{Chang J.}}},\ }\bibfield  {title} {\enquote {\bibinfo {title}
  {{Direct observation of orbital hybridisation in a cuprate
  superconductor}},}\ }\href {\doibase 10.1038/s41467-018-03266-0} {\bibfield
  {journal} {\bibinfo  {journal} {Nat. Commun.}\ }\textbf {\bibinfo {volume}
  {9}},\ \bibinfo {pages} {972} (\bibinfo {year} {2018})}\BibitemShut {NoStop}%
\bibitem [{\citenamefont {{J. Chang}}\ \emph {et~al.}(2013)\citenamefont {{J.
  Chang}}, \citenamefont {{M. M{\aa}nsson}}, \citenamefont {{S. Pailh{\`e}s}},
  \citenamefont {{T. Claesson}}, \citenamefont {{O. J. Lipscombe}},
  \citenamefont {{S. M. Hayden}}, \citenamefont {{L. Patthey}}, \citenamefont
  {{O. Tjernberg}},\ and\ \citenamefont {{J. Mesot}}}]{ChangNATC13}%
  \BibitemOpen
  \bibfield  {author} {\bibinfo {author} {\bibnamefont {{J. Chang}}}, \bibinfo
  {author} {\bibnamefont {{M. M{\aa}nsson}}}, \bibinfo {author} {\bibnamefont
  {{S. Pailh{\`e}s}}}, \bibinfo {author} {\bibnamefont {{T. Claesson}}},
  \bibinfo {author} {\bibnamefont {{O. J. Lipscombe}}}, \bibinfo {author}
  {\bibnamefont {{S. M. Hayden}}}, \bibinfo {author} {\bibnamefont {{L.
  Patthey}}}, \bibinfo {author} {\bibnamefont {{O. Tjernberg}}}, \ and\
  \bibinfo {author} {\bibnamefont {{J. Mesot}}},\ }\bibfield  {title} {\enquote
  {\bibinfo {title} {{Anisotropic breakdown of Fermi liquid quasiparticle
  excitations in overdoped La2-xSrxCuO4}},}\ }\href {\doibase
  10.1038/ncomms3559} {\bibfield  {journal} {\bibinfo  {journal} {Nat.
  Commun.}\ }\textbf {\bibinfo {volume} {4}},\ \bibinfo {pages} {2559}
  (\bibinfo {year} {2013})}\BibitemShut {NoStop}%
\bibitem [{\citenamefont {Chang}\ \emph {et~al.}(2008)\citenamefont {Chang},
  \citenamefont {Shi}, \citenamefont {Pailh\'es}, \citenamefont {M\aa{}nsson},
  \citenamefont {Claesson}, \citenamefont {Tjernberg}, \citenamefont
  {Bendounan}, \citenamefont {Sassa}, \citenamefont {Patthey}, \citenamefont
  {Momono}, \citenamefont {Oda}, \citenamefont {Ido}, \citenamefont {Guerrero},
  \citenamefont {Mudry},\ and\ \citenamefont {Mesot}}]{ChangPRB2008a}%
  \BibitemOpen
  \bibfield  {author} {\bibinfo {author} {\bibfnamefont {J.}~\bibnamefont
  {Chang}}, \bibinfo {author} {\bibfnamefont {M.}~\bibnamefont {Shi}}, \bibinfo
  {author} {\bibfnamefont {S.}~\bibnamefont {Pailh\'es}}, \bibinfo {author}
  {\bibfnamefont {M.}~\bibnamefont {M\aa{}nsson}}, \bibinfo {author}
  {\bibfnamefont {T.}~\bibnamefont {Claesson}}, \bibinfo {author}
  {\bibfnamefont {O.}~\bibnamefont {Tjernberg}}, \bibinfo {author}
  {\bibfnamefont {A.}~\bibnamefont {Bendounan}}, \bibinfo {author}
  {\bibfnamefont {Y.}~\bibnamefont {Sassa}}, \bibinfo {author} {\bibfnamefont
  {L.}~\bibnamefont {Patthey}}, \bibinfo {author} {\bibfnamefont
  {N.}~\bibnamefont {Momono}}, \bibinfo {author} {\bibfnamefont
  {M.}~\bibnamefont {Oda}}, \bibinfo {author} {\bibfnamefont {M.}~\bibnamefont
  {Ido}}, \bibinfo {author} {\bibfnamefont {S.}~\bibnamefont {Guerrero}},
  \bibinfo {author} {\bibfnamefont {C.}~\bibnamefont {Mudry}}, \ and\ \bibinfo
  {author} {\bibfnamefont {J.}~\bibnamefont {Mesot}},\ }\bibfield  {title}
  {\enquote {\bibinfo {title} {{Anisotropic quasiparticle scattering rates in
  slightly underdoped to optimally doped high-temperature
  ${\text{La}}_{2\ensuremath{-}x}{\text{Sr}}_{x}{\text{CuO}}_{4}$
  superconductors}},}\ }\href {\doibase 10.1103/PhysRevB.78.205103} {\bibfield
  {journal} {\bibinfo  {journal} {Phys. Rev. B}\ }\textbf {\bibinfo {volume}
  {78}},\ \bibinfo {pages} {205103} (\bibinfo {year} {2008})}\BibitemShut
  {NoStop}%
\bibitem [{\citenamefont {Sato}\ and\ \citenamefont
  {Naito}(1997)}]{SatoPhysicaC1997}%
  \BibitemOpen
  \bibfield  {author} {\bibinfo {author} {\bibfnamefont {H.}~\bibnamefont
  {Sato}}\ and\ \bibinfo {author} {\bibfnamefont {M.}~\bibnamefont {Naito}},\
  }\bibfield  {title} {\enquote {\bibinfo {title} {{Increase in the
  superconducting transition temperature by anisotropic strain effect in (001)
  La$_{1.85}$Sr$_{0.15}$CuO$_4$ thin films on LaSrAlO$_4$ substrates}},}\
  }\href {\doibase 10.1016/S0921-4534(96)00675-2} {\bibfield  {journal}
  {\bibinfo  {journal} {Phys. C}\ }\textbf {\bibinfo {volume} {274}},\ \bibinfo
  {pages} {221–226} (\bibinfo {year} {1997})}\BibitemShut {NoStop}%
\bibitem [{\citenamefont {Locquet}\ and\ \citenamefont
  {Williams}(1997)}]{LocquetActaP1997}%
  \BibitemOpen
  \bibfield  {author} {\bibinfo {author} {\bibfnamefont {J.}~\bibnamefont
  {Locquet}}\ and\ \bibinfo {author} {\bibfnamefont {E.}~\bibnamefont
  {Williams}},\ }\bibfield  {title} {\enquote {\bibinfo {title} {{Epitaxially
  Induced Defects in Sr- and O-doped La$_{2}$CuO$_{4}$ Thin Films Grown by MBE:
  Implications for Transport Properties}},}\ }\href {\doibase
  10.12693/APhysPolA.92.69} {\bibfield  {journal} {\bibinfo  {journal} {Acta
  Phys. Pol. A}\ }\textbf {\bibinfo {volume} {92}},\ \bibinfo {pages} {69–84}
  (\bibinfo {year} {1997})}\BibitemShut {NoStop}%
\bibitem [{\citenamefont {Nakamura}\ \emph {et~al.}(2000)\citenamefont
  {Nakamura}, \citenamefont {Goko}, \citenamefont {Hori}, \citenamefont {Uno},
  \citenamefont {Kikugawa},\ and\ \citenamefont {Fujita}}]{NakamuraPRB2000}%
  \BibitemOpen
  \bibfield  {author} {\bibinfo {author} {\bibfnamefont {Fumihiko}\
  \bibnamefont {Nakamura}}, \bibinfo {author} {\bibfnamefont {Tatsuo}\
  \bibnamefont {Goko}}, \bibinfo {author} {\bibfnamefont {Junya}\ \bibnamefont
  {Hori}}, \bibinfo {author} {\bibfnamefont {Yoshinori}\ \bibnamefont {Uno}},
  \bibinfo {author} {\bibfnamefont {Naoki}\ \bibnamefont {Kikugawa}}, \ and\
  \bibinfo {author} {\bibfnamefont {Toshizo}\ \bibnamefont {Fujita}},\
  }\bibfield  {title} {\enquote {\bibinfo {title} {{Role of two-dimensional
  electronic state in superconductivity in
  ${\mathrm{La}}_{2\ensuremath{-}x}{\mathrm{Sr}}_{x}{\mathrm{CuO}}_{4}$}},}\
  }\href {\doibase 10.1103/PhysRevB.61.107} {\bibfield  {journal} {\bibinfo
  {journal} {Phys. Rev. B}\ }\textbf {\bibinfo {volume} {61}},\ \bibinfo
  {pages} {107--110} (\bibinfo {year} {2000})}\BibitemShut {NoStop}%
\bibitem [{\citenamefont {Scalapino}(2012)}]{ScalapinoRMP12}%
  \BibitemOpen
  \bibfield  {author} {\bibinfo {author} {\bibfnamefont {D.~J.}\ \bibnamefont
  {Scalapino}},\ }\bibfield  {title} {\enquote {\bibinfo {title} {{A common
  thread: The pairing interaction for unconventional superconductors}},}\
  }\href {\doibase 10.1103/RevModPhys.84.1383} {\bibfield  {journal} {\bibinfo
  {journal} {Rev. Mod. Phys.}\ }\textbf {\bibinfo {volume} {84}},\ \bibinfo
  {pages} {1383--1417} (\bibinfo {year} {2012})}\BibitemShut {NoStop}%
\bibitem [{\citenamefont {Ofer}\ \emph {et~al.}(2006)\citenamefont {Ofer},
  \citenamefont {Bazalitsky}, \citenamefont {Kanigel}, \citenamefont {Keren},
  \citenamefont {Auerbach}, \citenamefont {Lord},\ and\ \citenamefont
  {Amato}}]{OferPRB2006}%
  \BibitemOpen
  \bibfield  {author} {\bibinfo {author} {\bibfnamefont {Rinat}\ \bibnamefont
  {Ofer}}, \bibinfo {author} {\bibfnamefont {Galina}\ \bibnamefont
  {Bazalitsky}}, \bibinfo {author} {\bibfnamefont {Amit}\ \bibnamefont
  {Kanigel}}, \bibinfo {author} {\bibfnamefont {Amit}\ \bibnamefont {Keren}},
  \bibinfo {author} {\bibfnamefont {Assa}\ \bibnamefont {Auerbach}}, \bibinfo
  {author} {\bibfnamefont {James~S.}\ \bibnamefont {Lord}}, \ and\ \bibinfo
  {author} {\bibfnamefont {Alex}\ \bibnamefont {Amato}},\ }\bibfield  {title}
  {\enquote {\bibinfo {title} {{Magnetic analog of the isotope effect in
  cuprates}},}\ }\href {\doibase 10.1103/PhysRevB.74.220508} {\bibfield
  {journal} {\bibinfo  {journal} {Phys. Rev. B}\ }\textbf {\bibinfo {volume}
  {74}},\ \bibinfo {pages} {220508} (\bibinfo {year} {2006})}\BibitemShut
  {NoStop}%
\bibitem [{\citenamefont {Ellis}\ \emph {et~al.}(2015)\citenamefont {Ellis},
  \citenamefont {Huang}, \citenamefont {Olalde-Velasco}, \citenamefont {Dantz},
  \citenamefont {Pelliciari}, \citenamefont {Drachuck}, \citenamefont {Ofer},
  \citenamefont {Bazalitsky}, \citenamefont {Berger}, \citenamefont {Schmitt},\
  and\ \citenamefont {Keren}}]{EllisPRB2015}%
  \BibitemOpen
  \bibfield  {author} {\bibinfo {author} {\bibfnamefont {David~Shai}\
  \bibnamefont {Ellis}}, \bibinfo {author} {\bibfnamefont {Yao-Bo}\
  \bibnamefont {Huang}}, \bibinfo {author} {\bibfnamefont {Paul}\ \bibnamefont
  {Olalde-Velasco}}, \bibinfo {author} {\bibfnamefont {Marcus}\ \bibnamefont
  {Dantz}}, \bibinfo {author} {\bibfnamefont {Jonathan}\ \bibnamefont
  {Pelliciari}}, \bibinfo {author} {\bibfnamefont {Gil}\ \bibnamefont
  {Drachuck}}, \bibinfo {author} {\bibfnamefont {Rinat}\ \bibnamefont {Ofer}},
  \bibinfo {author} {\bibfnamefont {Galina}\ \bibnamefont {Bazalitsky}},
  \bibinfo {author} {\bibfnamefont {Jorge}\ \bibnamefont {Berger}}, \bibinfo
  {author} {\bibfnamefont {Thorsten}\ \bibnamefont {Schmitt}}, \ and\ \bibinfo
  {author} {\bibfnamefont {Amit}\ \bibnamefont {Keren}},\ }\bibfield  {title}
  {\enquote {\bibinfo {title} {{Correlation of the superconducting critical
  temperature with spin and orbital excitations in
  $({\mathrm{Ca}}_{x}{\mathrm{La}}_{1\ensuremath{-}x})({\mathrm{Ba}}_{1.75\ensuremath{-}x}\text{La}$
  ${}_{0.25+x}){\mathrm{Cu}}_{3}{\mathrm{O}}_{y}$ as measured by resonant
  inelastic x-ray scattering}},}\ }\href {\doibase 10.1103/PhysRevB.92.104507}
  {\bibfield  {journal} {\bibinfo  {journal} {Phys. Rev. B}\ }\textbf {\bibinfo
  {volume} {92}},\ \bibinfo {pages} {104507} (\bibinfo {year}
  {2015})}\BibitemShut {NoStop}%
\bibitem [{\citenamefont {{Fratino L.}}\ \emph {et~al.}(2016)\citenamefont
  {{Fratino L.}}, \citenamefont {{Sémon P.}}, \citenamefont {{Sordi G.}},\
  and\ \citenamefont {{Tremblay A.-M. S.}}}]{FratinoSciRep2016}%
  \BibitemOpen
  \bibfield  {author} {\bibinfo {author} {\bibnamefont {{Fratino L.}}},
  \bibinfo {author} {\bibnamefont {{Sémon P.}}}, \bibinfo {author}
  {\bibnamefont {{Sordi G.}}}, \ and\ \bibinfo {author} {\bibnamefont
  {{Tremblay A.-M. S.}}},\ }\bibfield  {title} {\enquote {\bibinfo {title} {{An
  organizing principle for two-dimensional strongly correlated
  superconductivity}},}\ }\href {\doibase 10.1038/srep22715} {\bibfield
  {journal} {\bibinfo  {journal} {Sci. Rep.}\ }\textbf {\bibinfo {volume}
  {6}},\ \bibinfo {pages} {22715} (\bibinfo {year} {2016})}\BibitemShut
  {NoStop}%
\bibitem [{\citenamefont {Lichtensteiger}(2018)}]{LichtensteigerJAC2008}%
  \BibitemOpen
  \bibfield  {author} {\bibinfo {author} {\bibfnamefont {Céline}\ \bibnamefont
  {Lichtensteiger}},\ }\bibfield  {title} {\enquote {\bibinfo {title} {{{\it
  InteractiveXRDFit}: a new tool to simulate and fit X-ray diffractograms of
  oxide thin films and heterostructures}},}\ }\href {\doibase
  10.1107/S1600576718012840} {\bibfield  {journal} {\bibinfo  {journal}
  {Journal of Applied Crystallography}\ }\textbf {\bibinfo {volume} {51}},\
  \bibinfo {pages} {1745--1751} (\bibinfo {year} {2018})}\BibitemShut {NoStop}%
\bibitem [{\citenamefont {Ghiringhelli}\ \emph {et~al.}(2006)\citenamefont
  {Ghiringhelli}, \citenamefont {Piazzalunga}, \citenamefont {Dallera},
  \citenamefont {Trezzi}, \citenamefont {Braicovich}, \citenamefont {Schmitt},
  \citenamefont {Strocov}, \citenamefont {Betemps}, \citenamefont {Patthey},
  \citenamefont {Wang},\ and\ \citenamefont
  {Grioni}}]{ghiringhelliREVSCIINS2006}%
  \BibitemOpen
  \bibfield  {author} {\bibinfo {author} {\bibfnamefont {G.}~\bibnamefont
  {Ghiringhelli}}, \bibinfo {author} {\bibfnamefont {A.}~\bibnamefont
  {Piazzalunga}}, \bibinfo {author} {\bibfnamefont {C.}~\bibnamefont
  {Dallera}}, \bibinfo {author} {\bibfnamefont {G.}~\bibnamefont {Trezzi}},
  \bibinfo {author} {\bibfnamefont {L.}~\bibnamefont {Braicovich}}, \bibinfo
  {author} {\bibfnamefont {T.}~\bibnamefont {Schmitt}}, \bibinfo {author}
  {\bibfnamefont {V.~N.}\ \bibnamefont {Strocov}}, \bibinfo {author}
  {\bibfnamefont {R.}~\bibnamefont {Betemps}}, \bibinfo {author} {\bibfnamefont
  {L.}~\bibnamefont {Patthey}}, \bibinfo {author} {\bibfnamefont
  {X.}~\bibnamefont {Wang}}, \ and\ \bibinfo {author} {\bibfnamefont
  {M.}~\bibnamefont {Grioni}},\ }\bibfield  {title} {\enquote {\bibinfo {title}
  {{SAXES, a high resolution spectrometer for resonant x-ray emission in the
  400-1600eV energy range}},}\ }\href {\doibase 10.1063/1.2372731} {\bibfield
  {journal} {\bibinfo  {journal} {Rev. Sci. Instrum.}\ }\textbf {\bibinfo
  {volume} {77}},\ \bibinfo {eid} {113108} (\bibinfo {year}
  {2006})}\BibitemShut {NoStop}%
\bibitem [{\citenamefont {Strocov}\ \emph {et~al.}(2010)\citenamefont
  {Strocov}, \citenamefont {Schmitt}, \citenamefont {Flechsig}, \citenamefont
  {Schmidt}, \citenamefont {Imhof}, \citenamefont {Chen}, \citenamefont
  {Raabe}, \citenamefont {Betemps}, \citenamefont {Zimoch}, \citenamefont
  {Krempasky}, \citenamefont {Wang}, \citenamefont {Grioni},\ and\
  \citenamefont {Patthey}}]{strocovJSYNRAD2010}%
  \BibitemOpen
  \bibfield  {author} {\bibinfo {author} {\bibfnamefont {V.~N.}\ \bibnamefont
  {Strocov}}, \bibinfo {author} {\bibfnamefont {T.}~\bibnamefont {Schmitt}},
  \bibinfo {author} {\bibfnamefont {U.}~\bibnamefont {Flechsig}}, \bibinfo
  {author} {\bibfnamefont {T.}~\bibnamefont {Schmidt}}, \bibinfo {author}
  {\bibfnamefont {A.}~\bibnamefont {Imhof}}, \bibinfo {author} {\bibfnamefont
  {Q.}~\bibnamefont {Chen}}, \bibinfo {author} {\bibfnamefont {J.}~\bibnamefont
  {Raabe}}, \bibinfo {author} {\bibfnamefont {R.}~\bibnamefont {Betemps}},
  \bibinfo {author} {\bibfnamefont {D.}~\bibnamefont {Zimoch}}, \bibinfo
  {author} {\bibfnamefont {J.}~\bibnamefont {Krempasky}}, \bibinfo {author}
  {\bibfnamefont {X.}~\bibnamefont {Wang}}, \bibinfo {author} {\bibfnamefont
  {M.and Piazzalunga~A.}\ \bibnamefont {Grioni}}, \ and\ \bibinfo {author}
  {\bibfnamefont {L.}~\bibnamefont {Patthey}},\ }\bibfield  {title} {\enquote
  {\bibinfo {title} {{High-resolution soft X-ray beamline ADRESS at the Swiss
  Light Source for resonant inelastic X-ray scattering and angle-resolved
  photoelectron spectroscopies}},}\ }\href {\doibase 10.1107/S0909049510019862}
  {\bibfield  {journal} {\bibinfo  {journal} {J. Synchrotron Radiat.}\ }\textbf
  {\bibinfo {volume} {17}},\ \bibinfo {pages} {631--643} (\bibinfo {year}
  {2010})}\BibitemShut {NoStop}%
\bibitem [{\citenamefont {Radaelli}\ \emph {et~al.}(1994)\citenamefont
  {Radaelli}, \citenamefont {Hinks}, \citenamefont {Mitchell}, \citenamefont
  {Hunter}, \citenamefont {Wagner}, \citenamefont {Dabrowski}, \citenamefont
  {Vandervoort}, \citenamefont {Viswanathan},\ and\ \citenamefont
  {Jorgensen}}]{RadaelliPRB94}%
  \BibitemOpen
  \bibfield  {author} {\bibinfo {author} {\bibfnamefont {P.~G.}\ \bibnamefont
  {Radaelli}}, \bibinfo {author} {\bibfnamefont {D.~G.}\ \bibnamefont {Hinks}},
  \bibinfo {author} {\bibfnamefont {A.~W.}\ \bibnamefont {Mitchell}}, \bibinfo
  {author} {\bibfnamefont {B.~A.}\ \bibnamefont {Hunter}}, \bibinfo {author}
  {\bibfnamefont {J.~L.}\ \bibnamefont {Wagner}}, \bibinfo {author}
  {\bibfnamefont {B.}~\bibnamefont {Dabrowski}}, \bibinfo {author}
  {\bibfnamefont {K.~G.}\ \bibnamefont {Vandervoort}}, \bibinfo {author}
  {\bibfnamefont {H.~K.}\ \bibnamefont {Viswanathan}}, \ and\ \bibinfo {author}
  {\bibfnamefont {J.~D.}\ \bibnamefont {Jorgensen}},\ }\bibfield  {title}
  {\enquote {\bibinfo {title} {{Structural and superconducting properties of
  ${\mathrm{La}}_{2\mathrm{-}\mathit{x}}{\mathrm{Sr}}_{\mathit{x}}{\mathrm{CuO}}_{4}$
  as a function of Sr content}},}\ }\href {\doibase 10.1103/PhysRevB.49.4163}
  {\bibfield  {journal} {\bibinfo  {journal} {Phys. Rev. B}\ }\textbf {\bibinfo
  {volume} {49}},\ \bibinfo {pages} {4163--4175} (\bibinfo {year}
  {1994})}\BibitemShut {NoStop}%
\bibitem [{\citenamefont {Vasylechko}\ \emph {et~al.}(2000)\citenamefont
  {Vasylechko}, \citenamefont {Akselrud}, \citenamefont {Morgenroth},
  \citenamefont {Bismayer}, \citenamefont {Matkovskii},\ and\ \citenamefont
  {Savytskii}}]{VasylechkoJAC2000}%
  \BibitemOpen
  \bibfield  {author} {\bibinfo {author} {\bibfnamefont {L}~\bibnamefont
  {Vasylechko}}, \bibinfo {author} {\bibfnamefont {L}~\bibnamefont {Akselrud}},
  \bibinfo {author} {\bibfnamefont {W}~\bibnamefont {Morgenroth}}, \bibinfo
  {author} {\bibfnamefont {U}~\bibnamefont {Bismayer}}, \bibinfo {author}
  {\bibfnamefont {A}~\bibnamefont {Matkovskii}}, \ and\ \bibinfo {author}
  {\bibfnamefont {D}~\bibnamefont {Savytskii}},\ }\bibfield  {title} {\enquote
  {\bibinfo {title} {{The crystal structure of NdGaO3 at 100 K and 293 K based
  on synchrotron data}},}\ }\href {\doibase 10.1016/S0925-8388(99)00603-9}
  {\bibfield  {journal} {\bibinfo  {journal} {J. Alloy. Compd.}\ }\textbf
  {\bibinfo {volume} {297}},\ \bibinfo {pages} {46–52} (\bibinfo {year}
  {2000})}\BibitemShut {NoStop}%
\bibitem [{\citenamefont {Methfessel}\ \emph {et~al.}(2000)\citenamefont
  {Methfessel}, \citenamefont {van Schilfgaarde},\ and\ \citenamefont
  {Casali}}]{fplmto}%
  \BibitemOpen
  \bibfield  {author} {\bibinfo {author} {\bibfnamefont {M.}~\bibnamefont
  {Methfessel}}, \bibinfo {author} {\bibfnamefont {Mark}\ \bibnamefont {van
  Schilfgaarde}}, \ and\ \bibinfo {author} {\bibfnamefont {R.I.}\ \bibnamefont
  {Casali}},\ }\bibfield  {title} {\enquote {\bibinfo {title} {{A
  full-potential LMTO method based on smooth Hankel functions}},}\ }\href
  {\doibase 10.1007/3-540-46437-9_3} {\bibfield  {journal} {\bibinfo  {journal}
  {in Electronic Structure and Physical Properties of Solids: The Uses of the
  LMTO Method, Lecture Notes in Physics. H. Dreysse, ed.}\ }\textbf {\bibinfo
  {volume} {535}},\ \bibinfo {pages} {114–147} (\bibinfo {year}
  {2000})}\BibitemShut {NoStop}%
\bibitem [{\citenamefont {Marzari}\ \emph {et~al.}(2012)\citenamefont
  {Marzari}, \citenamefont {Mostofi}, \citenamefont {Yates}, \citenamefont
  {Souza},\ and\ \citenamefont {Vanderbilt}}]{MarzariRMP2012}%
  \BibitemOpen
  \bibfield  {author} {\bibinfo {author} {\bibfnamefont {Nicola}\ \bibnamefont
  {Marzari}}, \bibinfo {author} {\bibfnamefont {Arash~A.}\ \bibnamefont
  {Mostofi}}, \bibinfo {author} {\bibfnamefont {Jonathan~R.}\ \bibnamefont
  {Yates}}, \bibinfo {author} {\bibfnamefont {Ivo}\ \bibnamefont {Souza}}, \
  and\ \bibinfo {author} {\bibfnamefont {David}\ \bibnamefont {Vanderbilt}},\
  }\bibfield  {title} {\enquote {\bibinfo {title} {{Maximally localized Wannier
  functions: Theory and applications}},}\ }\href {\doibase
  10.1103/RevModPhys.84.1419} {\bibfield  {journal} {\bibinfo  {journal} {Rev.
  Mod. Phys.}\ }\textbf {\bibinfo {volume} {84}},\ \bibinfo {pages}
  {1419--1475} (\bibinfo {year} {2012})}\BibitemShut {NoStop}%
\bibitem [{\citenamefont {Aryasetiawan}\ \emph {et~al.}(2004)\citenamefont
  {Aryasetiawan}, \citenamefont {Imada}, \citenamefont {Georges}, \citenamefont
  {Kotliar}, \citenamefont {Biermann},\ and\ \citenamefont
  {Lichtenstein}}]{AryasetiawanPRB2004}%
  \BibitemOpen
  \bibfield  {author} {\bibinfo {author} {\bibfnamefont {F.}~\bibnamefont
  {Aryasetiawan}}, \bibinfo {author} {\bibfnamefont {M.}~\bibnamefont {Imada}},
  \bibinfo {author} {\bibfnamefont {A.}~\bibnamefont {Georges}}, \bibinfo
  {author} {\bibfnamefont {G.}~\bibnamefont {Kotliar}}, \bibinfo {author}
  {\bibfnamefont {S.}~\bibnamefont {Biermann}}, \ and\ \bibinfo {author}
  {\bibfnamefont {A.~I.}\ \bibnamefont {Lichtenstein}},\ }\bibfield  {title}
  {\enquote {\bibinfo {title} {{Frequency-dependent local interactions and
  low-energy effective models from electronic structure calculations}},}\
  }\href {\doibase 10.1103/PhysRevB.70.195104} {\bibfield  {journal} {\bibinfo
  {journal} {Phys. Rev. B}\ }\textbf {\bibinfo {volume} {70}},\ \bibinfo
  {pages} {195104} (\bibinfo {year} {2004})}\BibitemShut {NoStop}%
\bibitem [{\citenamefont {Miyake}\ and\ \citenamefont
  {Aryasetiawan}(2008)}]{MiyakePRB2008}%
  \BibitemOpen
  \bibfield  {author} {\bibinfo {author} {\bibfnamefont {Takashi}\ \bibnamefont
  {Miyake}}\ and\ \bibinfo {author} {\bibfnamefont {F.}~\bibnamefont
  {Aryasetiawan}},\ }\bibfield  {title} {\enquote {\bibinfo {title} {{Screened
  Coulomb interaction in the maximally localized Wannier basis}},}\ }\href
  {\doibase 10.1103/PhysRevB.77.085122} {\bibfield  {journal} {\bibinfo
  {journal} {Phys. Rev. B}\ }\textbf {\bibinfo {volume} {77}},\ \bibinfo
  {pages} {085122} (\bibinfo {year} {2008})}\BibitemShut {NoStop}%
\bibitem [{\citenamefont {Miyake}\ \emph {et~al.}(2009)\citenamefont {Miyake},
  \citenamefont {Aryasetiawan},\ and\ \citenamefont {Imada}}]{MiyakePRB2009}%
  \BibitemOpen
  \bibfield  {author} {\bibinfo {author} {\bibfnamefont {Takashi}\ \bibnamefont
  {Miyake}}, \bibinfo {author} {\bibfnamefont {Ferdi}\ \bibnamefont
  {Aryasetiawan}}, \ and\ \bibinfo {author} {\bibfnamefont {Masatoshi}\
  \bibnamefont {Imada}},\ }\bibfield  {title} {\enquote {\bibinfo {title} {{Ab
  initio procedure for constructing effective models of correlated materials
  with entangled band structure}},}\ }\href {\doibase
  10.1103/PhysRevB.80.155134} {\bibfield  {journal} {\bibinfo  {journal} {Phys.
  Rev. B}\ }\textbf {\bibinfo {volume} {80}},\ \bibinfo {pages} {155134}
  (\bibinfo {year} {2009})}\BibitemShut {NoStop}%
\bibitem [{\citenamefont {Locquet}\ \emph {et~al.}(1998)\citenamefont
  {Locquet}, \citenamefont {Perret}, \citenamefont {Fompeyrine}, \citenamefont
  {Machler}, \citenamefont {Seo},\ and\ \citenamefont {{Van
  Tendeloo}}}]{LocquetNat98}%
  \BibitemOpen
  \bibfield  {author} {\bibinfo {author} {\bibfnamefont {J.~P.}\ \bibnamefont
  {Locquet}}, \bibinfo {author} {\bibfnamefont {J.}~\bibnamefont {Perret}},
  \bibinfo {author} {\bibfnamefont {J.}~\bibnamefont {Fompeyrine}}, \bibinfo
  {author} {\bibfnamefont {E.}~\bibnamefont {Machler}}, \bibinfo {author}
  {\bibfnamefont {J.~W.}\ \bibnamefont {Seo}}, \ and\ \bibinfo {author}
  {\bibfnamefont {G.}~\bibnamefont {{Van Tendeloo}}},\ }\bibfield  {title}
  {\enquote {\bibinfo {title} {{Doubling the critical temperature of
  La$_{1.9}$Sr$_{0.1}$CuO$_4$ using epitaxial strain}},}\ }\href
  {http://dx.doi.org/10.1038/28810} {\bibfield  {journal} {\bibinfo  {journal}
  {Nature}\ }\textbf {\bibinfo {volume} {394}},\ \bibinfo {pages} {453--456}
  (\bibinfo {year} {1998})}\BibitemShut {NoStop}%
\bibitem [{\citenamefont {Takagi}\ \emph {et~al.}(1989)\citenamefont {Takagi},
  \citenamefont {Ido}, \citenamefont {Ishibashi}, \citenamefont {Uota},
  \citenamefont {Uchida},\ and\ \citenamefont {Tokura}}]{TakagiPRB1989}%
  \BibitemOpen
  \bibfield  {author} {\bibinfo {author} {\bibfnamefont {H.}~\bibnamefont
  {Takagi}}, \bibinfo {author} {\bibfnamefont {T.}~\bibnamefont {Ido}},
  \bibinfo {author} {\bibfnamefont {S.}~\bibnamefont {Ishibashi}}, \bibinfo
  {author} {\bibfnamefont {M.}~\bibnamefont {Uota}}, \bibinfo {author}
  {\bibfnamefont {S.}~\bibnamefont {Uchida}}, \ and\ \bibinfo {author}
  {\bibfnamefont {Y.}~\bibnamefont {Tokura}},\ }\bibfield  {title} {\enquote
  {\bibinfo {title} {{Superconductor-to-nonsuperconductor transition in
  (La$_{1-\mathrm{x}}$Sr$_{\mathrm{x}}$)$_{2}$CuO$_{4}$ as investigated by
  transport and magnetic measurements}},}\ }\href {\doibase
  10.1103/PhysRevB.40.2254} {\bibfield  {journal} {\bibinfo  {journal} {Phys.
  Rev. B}\ }\textbf {\bibinfo {volume} {40}},\ \bibinfo {pages} {2254--2261}
  (\bibinfo {year} {1989})}\BibitemShut {NoStop}%
\end{thebibliography}%

\clearpage
\beginsupplement
\onecolumngrid

\section{Supplementary Figures}

\begin{figure}[b]
 	\begin{center}
  		\includegraphics[height=0.7\textheight]{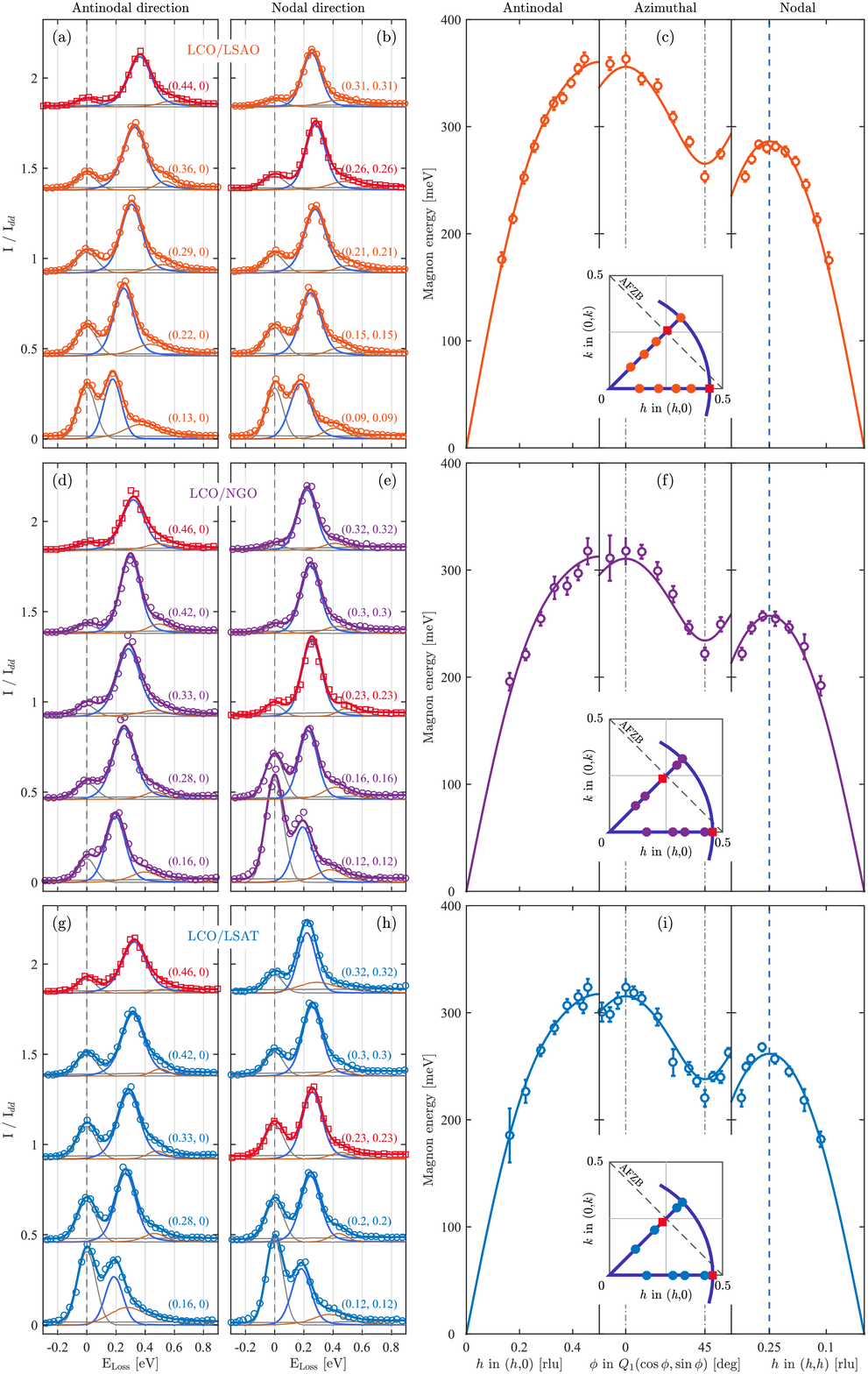}
 	\end{center}
 	\caption{\textbf{Magnon Dispersion of thin films.} In (a) and (b) are presented raw spectra of LCO/LSAO along with the fits (and its single components) for antinodal and nodal directions respectively. Similar data are presented for LCO/NGO in (d) and (e) and for LCO/LSAT in (g) and (h). Each spectra is at the \textbf{q} vector as indicated, which is also schematically illustrated in the respective insets. Magnon dispersions and the respective Hubbard model fits are presented for LCO/LSAO, LCO/NGO and LCO/LSAT in (c), (f) and (i) respectively, for direction as indicated. The error bars are three times the standard deviation obtained from the fits. In (c), (f) and (i) $Q_1$ takes different values for each compound due to different incident energies and in-plane lattice parameters, resulting in $0.4437$ for LCO$\slash$LSAO, $0.4564$ for LCO$\slash$NGO and $0.4568$ for LCO$\slash$LSAT. Source data are provided as a Source Data file.}\label{fig:S1}
\end{figure}

\begin{figure}[htb]
 	\begin{center}
 		\includegraphics[width=0.666\textwidth]{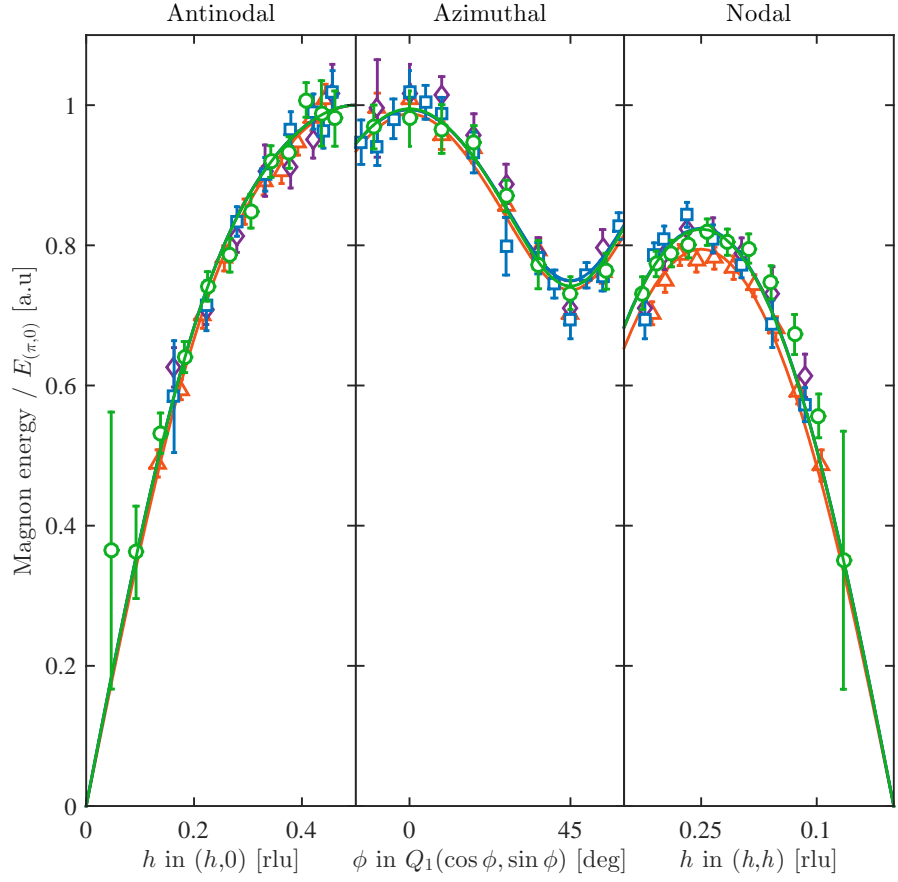}
 	\end{center}
 	\caption{\textbf{Scaling the Magnon Dispersions.} Dispersion of the magnetic excitations for all the measured samples (LCO$\slash$LSAO, LCO$\slash$NGO, LCO$\slash$LSAT and LCO$\slash$STO) scaled to the maximum along the antinodal direction $E_\mathrm{(\pi,0)}$. The error bars are three times the standard deviation obtained from the fits. Source data are provided as a Source Data file.}\label{fig:S3}
\end{figure}

\begin{figure}[hb]
 	\begin{center}
 		\includegraphics[width=0.7\textwidth]{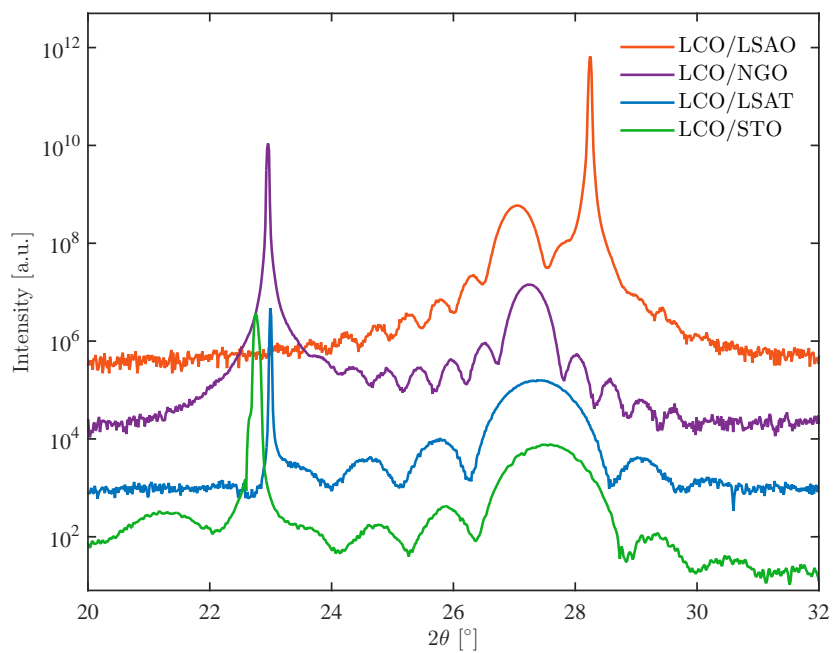}
 	\end{center}
 	\caption{\textbf{2$\theta$ scans of thin films.} X-ray diffraction measurements allowing the extraction of the $c$ lattice parameter (main peak around $27^\circ-28^\circ$) and the thickness for each sample as indicated. Sharp intense peaks around $23^\circ$ and $28^\circ$ belong to the substrates. Incident $1.5406$~\AA\ x rays were used for these measurements. Source data are provided as a Source Data file.}\label{fig:S2}
\end{figure}

\begin{figure}[htb]
 	\begin{center}
 		\includegraphics[width=0.7\textwidth]{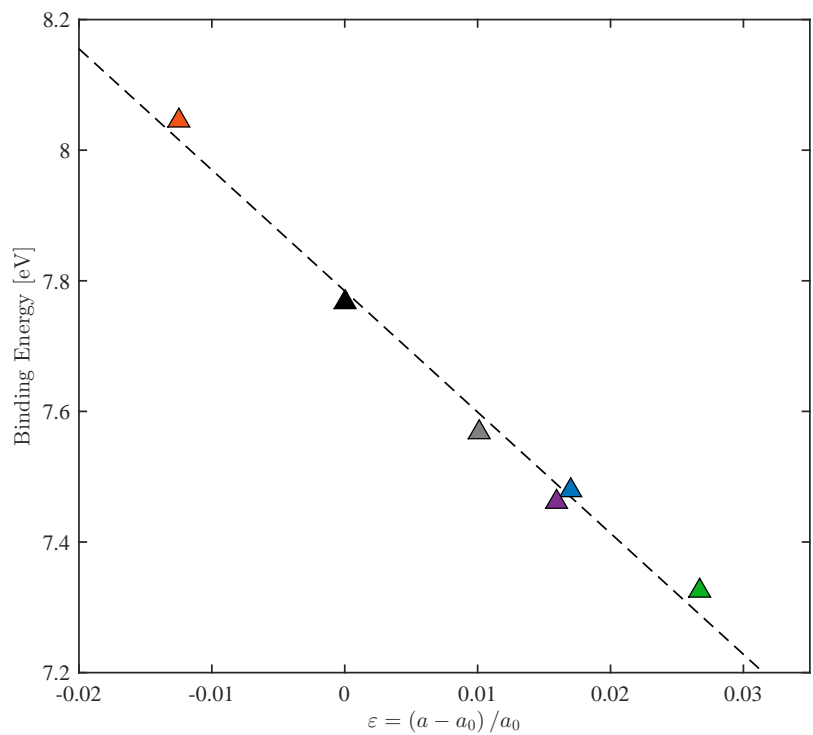}
 	\end{center}
 	\caption{\textbf{Evolution of the oxygen $p$ bands versus strain.} Binding energies of the lower edge of the oxygen $p$ bands (with $p_x/p_y$ character), obtained from DFT calculations, as a function of strain $\varepsilon$. Source data are provided as a Source Data file.}\label{fig:S4}
\end{figure}

\clearpage
\section{Supplementary Table}

\begin{table}[thb]
\begin{center}
\begin{ruledtabular}
    \caption{\textbf{Structural thin-film parameters \it{vs.} $T_\mathrm{c}$.} Lattice parameters ($a$ and $c$) and superconducting transition temperature $T_\mathrm{c}$ for optimally doped \LSCO{} thin films grown on substrates as indicated, as a function of thickness h. In-plane lattice parameters $a_\mathrm{S}$ of the respective substrates are indicated for completeness. Bulk \LSCO{} is also presented for the dopings $x$ considered for the thin films. All data are extracted from references as indicated in the last column. Source data are provided as a Source Data file.}\label{tab:tab2}
\begin{tabular}{cccccccc}
Substrate & h [nm] & Doping [$x$] & $a_\mathrm{S}$ [\AA] & $a$ [\AA] & $c$ [\AA] & $T_\mathrm{c}$ [K] & Ref.\\ \hline
SrTiO$_3$ & 15 & 0.10 & 3.905 & 3.80 & 13.17 & 10 & \cite{LocquetNat98} \\
SrTiO$_3$ & 50 & 0.15 & 3.905 & $\sim$~3.870 & $\sim$~13.16 & $\sim$~23.9 & \cite{SatoPhysicaC1997} \\
SrTiO$_3$ & 50 & $\sim$~0.16 & 3.905 & --- & $\sim$~13.18 & $\sim$~23 & \cite{LocquetActaP1997} \\
SrTiO$_3$ & 200 & $\sim$~0.16 & 3.905 & --- & $\sim$~13.20 & $\sim$~28 & \cite{LocquetActaP1997} \\
SrTiO$_3$ & 200 & 0.15 & 3.905 & 3.837 & 13.18 & 27.4 & \cite{SatoPhysicaC1997} \\
NdGaO$_3$ & 50 & 0.15 & 3.842 & $\sim$~3.797 & $\sim$~13.13 & $\sim$~18.2 & \cite{SatoPhysicaC1997} \\
LaSrAlO$_4$ & 15 & 0.10 & 3.754 & 3.76 & 13.31 & 49.1 & \cite{LocquetNat98} \\
LaSrAlO$_4$ & 50 & 0.15 & 3.756 & $\sim$~3.756 & $\sim$~13.26 & $\sim$~40.7 & \cite{SatoPhysicaC1997} \\
LaSrAlO$_4$ & 50 & $\sim$~0.15 & 3.756 & --- & 13.29 & $\sim$~38 & \cite{LocquetActaP1997} \\
LaSrAlO$_4$ & 200 & 0.15 & 3.7564 & 3.762 & 13.29 & 43.8 & \cite{SatoPhysicaC1997} \\ \hline
Bulk \\ \hline
--- & --- & 0.10 & --- & 3.778 & 13.21 & 27 & \cite{TakagiPRB1989} \\
--- & --- & 0.15 & --- & 3.777 & 13.23 & 36.5 & \cite{TakagiPRB1989} \\
\end{tabular}
\end{ruledtabular}
\end{center}
\end{table}

\end{document}